\documentclass[12pt]{article}
\usepackage[letterpaper, margin=1in]{geometry}
\usepackage{amsmath, amssymb, amscd, amsthm, amsfonts}
\usepackage{graphicx}
\usepackage{setspace}
\usepackage{needspace}

\usepackage{subcaption}
\usepackage{here}
\usepackage{enumitem}
\usepackage[affil-it]{authblk}

\usepackage[utf8]{inputenc}
\usepackage{booktabs}

\usepackage{natbib}
\usepackage{bm}

\usepackage{hyperref}

\usepackage{threeparttable}

\usepackage{docmute}

\renewcommand{\topfraction}{0.90}
\renewcommand{\bottomfraction}{0.80}
\renewcommand{\textfraction}{0.12}
\renewcommand{\floatpagefraction}{0.80}
\makeatletter
\renewcommand\section{\@startsection{section}{1}{\z@}{-3ex \@plus -1ex \@minus -.2ex}{1.8ex \@plus .2ex}{\normalfont\Large\bfseries}}
\renewcommand\subsection{\@startsection{subsection}{2}{\z@}{-2.8ex \@plus -1ex \@minus -.2ex}{1.5ex \@plus .2ex}{\normalfont\large\bfseries}}
\makeatother

\newif\ifhidetext
\hidetextfalse 
\newcommand{\hidetext}[1]{%
  \ifhidetext
    [hidden for blinding]
  \else
    #1
  \fi
}

\newcommand{\hideauthor}[1]{%
  \ifhidetext
    \author{}
  \else
    #1
  \fi
}

\def\spacingset#1{\renewcommand{\baselinestretch}%
{#1}\small\normalsize} \spacingset{1}

\title{Two-sample Comparison through Additive Tree Models for Density Ratios}
\hideauthor{
\author[1]{Naoki Awaya}
\author[2]{Yuliang Xu}
\author[2]{Li Ma}
\affil[1]{School of Political Science and Economics, Waseda University}
\affil[2]{Department of Statistics \& Data Science Institute, University of Chicago}
}
\date{}

\usepackage{xcolor}
\newcommand\rnew[1]{#1}
\newcommand\yxnew[1]{#1}
\newcommand\nanew[1]{#1}
\newcommand\yxcm[1]{}

\newcommand{\argmax}{\mathop{\rm arg~max}\limits}
\newcommand{\argmin}{\mathop{\rm arg~min}\limits}
\newcommand\real{\mathbb{R}}

\newtheorem{cor}{Corollary}
\newtheorem{prop}{Proposition}

\usepackage{amsmath}
  {
      \theoremstyle{plain}
      
  }

\newcommand{\1}{\mathbf{1}}

\newcommand{\x}{\mathbf{x}}
\newcommand{\y}{\mathbf{y}}
\newcommand{\h}{\mathbf{h}}

\newcommand{\Q}{\mathbf{Q}}

\newcommand{\C}{\mathcal{C}}
\newcommand{\A}{\mathcal{A}}
\newcommand{\B}{\mathcal{B}}
\newcommand{\T}{\mathcal{T}}
\newcommand{\Z}{\mathbb{Z}}
\newcommand{\Polya}{P\'{o}lya\ }
\newcommand{\F}{\mathbf{F}}
\newcommand{\tF}{\tilde{\mathbf{F}}}
\newcommand{\G}{\mathbf{G}}
\newcommand{\tG}{\tilde{\mathbf{G}}}
\newcommand{\cP}{\mathcal{P}}

\newcommand{\D}{\mathbf{D}}

\newcommand{\R}{\mathbb{R}}
\newcommand{\E}{\mathbb{E}}

\newcommand{\Var}{\mathrm{Var}}
\newcommand{\Cov}{\mathrm{Cov}}

\newcommand{\Bspace}{\mathcal{B}^2}

\begin{document}

\documentclass[12pt]{article}
\usepackage[letterpaper, margin=1in]{geometry}
\usepackage{amsmath, amssymb, amscd, amsthm, amsfonts}
\usepackage{graphicx}
\usepackage{setspace}
\usepackage{needspace}

\usepackage{subcaption}
\usepackage{here}
\usepackage{enumitem}
\usepackage[affil-it]{authblk}

\usepackage[utf8]{inputenc}
\usepackage{booktabs}

\usepackage{natbib}
\usepackage{bm}

\usepackage{xr-hyper}
\usepackage{hyperref}
\externaldocument[][nocite]{JASA_submission_R2/supplement_JASA_Aug10_2026}
  [supplement_JASA_Aug10_2026.pdf]

\usepackage{threeparttable} 

\setcounter{topnumber}{2}
\setcounter{bottomnumber}{1}
\setcounter{totalnumber}{3}
\renewcommand{\topfraction}{0.90}
\renewcommand{\bottomfraction}{0.80}
\renewcommand{\textfraction}{0.12}
\renewcommand{\floatpagefraction}{0.80}
\setlength{\textfloatsep}{12pt plus 2pt minus 2pt}
\setlength{\floatsep}{10pt plus 2pt minus 2pt}
\setlength{\intextsep}{10pt plus 2pt minus 2pt}
\captionsetup{skip=6pt}
\makeatletter
\renewcommand\section{\@startsection{section}{1}{\z@}{-3ex \@plus -1ex \@minus -.2ex}{1.8ex \@plus .2ex}{\normalfont\Large\bfseries}}
\renewcommand\subsection{\@startsection{subsection}{2}{\z@}{-2.8ex \@plus -1ex \@minus -.2ex}{1.5ex \@plus .2ex}{\normalfont\large\bfseries}}
\makeatother

\newif\ifhidetext
\hidetextfalse 
\newcommand{\hidetext}[1]{%
  \ifhidetext
    [hidden for blinding]
  \else
    #1
  \fi
}

\newcommand{\hideauthor}[1]{%
  \ifhidetext
    \author{}
  \else
    #1
  \fi
}

\def\spacingset#1{\renewcommand{\baselinestretch}%
{#1}\small\normalsize} \spacingset{1}

\oddsidemargin 0pt
\evensidemargin 0pt
\marginparwidth 40pt
\marginparsep 10pt
\topmargin -20pt
\headsep 10pt
\textheight 8.7in
\textwidth 6.65in
\linespread{1}

\title{Two-sample Comparison through Additive Tree Models for Density Ratios}
\hideauthor{
\author[1]{Naoki Awaya}
\author[2]{Yuliang Xu}
\author[2]{Li Ma}
\affil[1]{School of Political Science and Economics, Waseda University}
\affil[2]{Department of Statistics \& Data Science Institute, University of Chicago}
}
\date{}

\usepackage{xcolor}
\newcommand\yxnew[1]{{\color{blue} {#1}}} 
\newcommand\yxcm[1]{{\color{red} {(YX: #1)}}} 

\newcommand\nanew[1]{#1}
\newcommand\rnew[1]{{\color{blue} {#1}}}

\newcommand{\argmax}{\mathop{\rm arg~max}\limits}
\newcommand{\argmin}{\mathop{\rm arg~min}\limits}
\newcommand\real{\mathbb{R}}

\newtheorem{thm}{Theorem}
\newtheorem{lem}{Lemma}
\newtheorem{cor}{Corollary}
\newtheorem{prop}{Proposition}

\usepackage{amsmath}               
  {
      \theoremstyle{plain}
      \newtheorem{assumption}{Assumption}
  }

\newcommand{\1}{\mathbf{1}}

\newcommand{\x}{\mathbf{x}}
\newcommand{\y}{\mathbf{y}}
\newcommand{\h}{\mathbf{h}}

\newcommand{\Q}{\mathbf{Q}}

\newcommand{\C}{\mathcal{C}}
\newcommand{\A}{\mathcal{A}}
\newcommand{\B}{\mathcal{B}}
\newcommand{\T}{\mathcal{T}}
\newcommand{\Z}{\mathbb{Z}}
\newcommand{\Polya}{P\'{o}lya\ }
\newcommand{\F}{\mathbf{F}}
\newcommand{\tF}{\tilde{\mathbf{F}}}
\newcommand{\G}{\mathbf{G}}
\newcommand{\tG}{\tilde{\mathbf{G}}}
\newcommand{\cP}{\mathcal{P}}

\newcommand{\D}{\mathbf{D}}

\newcommand{\R}{\mathbb{R}}
\newcommand{\E}{\mathbb{E}}

\newcommand{\Var}{\mathrm{Var}}
\newcommand{\Cov}{\mathrm{Cov}}

\newcommand{\Bspace}{\mathcal{B}^2}

\begin{document}
\maketitle

\begin{abstract}
The ratio of two densities provides a direct characterization of their differences. 
We consider the two-sample comparison problem by estimating this ratio given i.i.d.\ observations from two distributions.
To this end, we propose additive tree models for density ratio estimation along with efficient algorithms using a new loss function, the balancing loss. 
The loss allows tree-based models to be trained using several algorithms originally designed for supervised learning, such as forward-stagewise optimization and gradient boosting. 
Moreover, 
the balancing loss resembles 
an exponential family kernel, and it can serve as a pseudo-likelihood with conjugate priors.
This property enables generalized Bayesian inference on the density ratio using backfitting samplers designed for Bayesian additive regression trees (BART). 
Our Bayesian strategy provides uncertainty quantification for the inferred density ratio, which is critical for applications involving high-dimensional and data-limited distributions with potentially substantial uncertainty. 
We further show connections of the balancing loss to the exponential loss in binary classification and to the variational form of $f$-divergence, particularly the squared Hellinger distance. 
Numerical experiments demonstrate that the proposed boosting algorithms achieve high accuracy at low computational cost, and that the Bayesian additive tree is competitive in accuracy while additionally providing uncertainty quantification, which the compared estimators do not offer.
Finally, we demonstrate its application to assessing the quality of generative models for microbiome compositional data.
\end{abstract}


\newpage
\doublespacing
\vspace{-1em}
\section{Introduction}
\vspace{-1em}
\rnew{Many applications compare two groups of observations to assess their differences.} The classical formulation of the two-sample comparison often involves parametric or semiparametric assumptions on the underlying distributions and focuses on testing the null hypothesis of no difference with or without a given alternative in mind. However, this hypothesis testing perspective is often inadequate for most modern applications of two-sample comparison. \rnew{Practitioners typically also seek to characterize the underlying difference.} This problem lies at the core of numerous other modern statistical analyses and machine learning objectives.

For example, in large-scale biomedical studies involving the comparison of genetic measurements and various biomarkers between patients and controls, there is often no question that some types of two-sample differences do exist, and with a sufficiently large sample size, the null of no difference at all {\em will} be rejected. It is more helpful to specifically pinpoint what such differences are in order to help understand their biological implication. 
Another timely application is the assessment of generative models. 
The key to assessing the quality of different models and algorithms for generating synthetic samples that imitate the real sampling distribution of the training data lies in deciphering the distributional difference between the real sampling distribution and that induced by the generative models, which is essentially a two-sample comparison problem.
Some generative models, such as generative adversarial networks (GANs) \citep{goodfellow2014generative}, essentially iteratively apply two-sample comparison and use the estimated misfit between the two distributions to  sequentially improve the generative model. Two-sample comparison is also a quintessential problem in causal inference, in which consistent estimation of treatment effects often relies on correctly adjusting for the difference in the covariate distributions between the treated and the control groups. 

\rnew{We focus on nonparametric two-sample comparison through estimation rather than hypothesis testing, using} the density ratio function as the functional ``statistic'' of interest for characterizing two-sample differences. The density ratio fully summarizes how two distributions differ across the sample space, \rnew{and thus supports both hypothesis testing and characterization of the difference.} If, for example, the estimated density ratio deviates ``significantly'' away from 1 on parts of the sample space, it allows us to draw statistical conclusions regarding whether, where, and how the two distributions differ. At the same time, the notion of ``significant'' deviation will require some form of uncertainty quantification for the inferred density ratio, which will become available under our formulation.

One may reasonably raise the concern that nonparametric density ratio learning is simply ``too hard'' for complex, high-dimensional data because density estimation itself in such settings is already a challenging problem. Why then would one attempt an even harder problem of learning the ratio of two densities?
Though it might appear counterintuitive, we argue that density ratio estimation (DRE) is in fact a much easier problem than density estimation in most modern applications. A density by definition is always a density ratio, with the denominator being the corresponding base measure against which the density is computed. The typical base measure, for example, the Lebesgue measure or the uniform density in continuous cases, is generally chosen out of mathematical convenience and lack of prior knowledge, without taking any account of the dataset at hand (other than that the observations are continuous). This base measure generally does not resemble any sampling distribution in high-dimensional or otherwise complex problems. When good reliable knowledge of the sampling distribution does exist, one would be in the much simpler classical setting of a ``goodness-of-fit'' problem. In contrast, in most applications of two-sample comparison, the two distributions are often quite similar to each other and can serve as a ``baseline of reference'' for each other. As such, the density ratio between these two samples then is often much simpler than the underlying densities (i.e., their ratios with respect to the Lebesgue measure). 

We believe this explains a well-known observation in the DRE literature, that though estimating a density ratio can be done by first separately estimating the two densities and then computing their ratios based on the density estimates, it is usually more effective to adopt a method directly aimed at the density ratio, especially when the difference exists only in a small subset of the sample space or in a subset of the dimensions. Several authors have adopted this perspective and proposed methods directly targeting the density ratio. Some approaches include kernel-based methods \citep[e.g.][]{sugiyama2007direct, kanamori2009least, hido2011statistical, liu2013change, yamada2013relative} and more recently neural network-based methods \citep{kato2021non}. 

\rnew{Beyond the stand-alone two-sample comparison considered here, DRE is used in learning methods where two-sample comparison arises directly or indirectly as a subproblem. Examples include}
covariate shift correction \citep{shimodaira2000improving, sugiyama2007covariate,reddi2015doubly, stojanov2019low}, outlier detection \citep{hido2011statistical, kato2021non}, causal inference \citep{uehara2020off}, and change point detection \citep{kawahara2009change, liu2013change, du2015online}.
Additionally, estimating density ratios has been utilized in Markov Chain Monte Carlo (MCMC) for intractable likelihoods \citep{kaji2023metropolis}.

We list several desiderata for learning the density ratio and discuss how we aim to achieve them. 
First, the model or function class representing the density ratio should be flexible enough to effectively approximate complex forms of the underlying truth. 
At the same time, it should scale well with both sample size and dimensionality.
To these ends, we propose a class of density ratio learners based on additive tree ensembles \citep[see e.g.,][]{friedman2000additive,hastie2009elements}. 
Additive tree ensembles are a natural candidate for the density ratio problem, because of their success in classification problems. In fact, in the machine learning literature, a widely used technique to learn the density ratio between two samples is the so-called ``density-ratio trick'' \citep{sugiyama2012density, durkan2020contrastive, thomas2022likelihood}, which involves first fitting a binary classifier to the two samples, and then, based on the estimated posterior odds of the sample label, computing the density ratio. Additive tree boosting such as AdaBoost \citep{friedman2000additive} is among the most effective methods for binary classification.

In contrast to this popular perspective, we hold that DRE should not and does not need to involve the indirect route of inverting a binary classifier based on additive tree ensembles, but can instead be directly learned from the latter. 
\nanew{
    To establish the direct route, we propose a new loss function called ``balancing loss'', which has a functional form similar to the exponential loss used in the classification AdaBoost \citep{friedman2000additive} but is minimized by  the square root of the true density ratio.
 \rnew{In our experiments, the resulting estimates are more accurate than those of the density-ratio trick with AdaBoost, which uses the same class of additive tree ensembles, in nearly all settings, and the advantage is clearest when the sample sizes are unbalanced, that is, when the setting is more challenging for classifiers that aim to maximize the total successful classification rate.}
    We demonstrate the limitation of adopting the classification AdaBoost in the experiments. 
}

In terms of computational strategies, we introduce two boosting algorithms for estimating the additive tree model on density ratios. 
We use an additive tree ensemble as the function class for approximating the density ratio, by adding tree-based weak learners.  
The first algorithm is based on forward-stagewise fitting \citep{hastie2009elements} and it operationally resembles AdaBoost by minimizing the balancing loss by repeatedly fitting a single-tree weak learner to reweighted observations.
The second algorithm follows the general paradigm of gradient boosting \citep{friedman2001greedy}.

These algorithms inherit important advantages of supervised tree boosting --- \rnew{they achieve good numerical performance at low computational cost}, especially when the difference lies in a marginal distribution of a subset of the variables.
\rnew{A small learning rate regularizes each base learner's contribution to avoid overfitting}, following the standard approach in supervised boosting \citep{hastie2009elements, friedman2001greedy}.

\rnew{The second desideratum is uncertainty quantification for the estimated density ratio to support statistical inference.} \rnew{Most existing DRE methods, including the kernel-based estimators and the density-ratio trick with boosted classifiers, produce only point estimates. Bayesian nonparametric models targeting the density ratio directly have been developed using a single tree model}\citep{ma2011coupling,soriano2017probabilistic}. While they provide a means to examine posterior uncertainty in a fully Bayesian manner, piecewise-constant functions represented by single trees of controlled complexity are very restrictive and therefore inadequate for approximating complex density ratios.
\nanew{Also, more importantly, they introduce generative models for the two samples, and this approach can be inefficient in terms of the two-sample comparison, since estimating the common distributional structure is redundant.
\rnew{Generalizing these Bayesian models to additive trees focused on the two-sample difference is therefore challenging.}
}
To overcome this, we propose 
a generalized Bayesian approach \citep{bissiri2016general} using the balancing loss to construct a pseudo-likelihood for this model class.  
Interestingly, we show that, due to the resemblance of the balancing loss to an exponential family kernel, there exists a natural conjugate prior for the model parameters under the pseudo-likelihood, which allows standard sampling strategies for BART models \citep{chipman2010bart} to be directly adopted for sampling the Gibbs posterior on the density ratio.

The remainder of the paper is organized as follows. Section~2 formulates the balancing loss and introduces the boosting algorithms. 
Section 3 proposes the generalized Bayesian inference based on the balancing loss and the BART model designed for density ratio estimation. 
Sections 4 and 5 present numerical experiments and a microbiome application, and Section 6 concludes.


\vspace{-2em}
\section{\nanew{Boosting Algorithms for Density Ratio Estimation}}
\vspace{-0.5em}
\nanew{In this section, we propose new boosting algorithms for density ratio estimation with a novel loss function called balancing loss.
We first formulate the balancing loss and discuss its connection to other learning problems.
Then, we introduce optimization procedures minimizing the loss, using the boosting algorithms \citep{friedman2001greedy}.
}
\vspace{-1.5em}
\subsection{The Balancing Loss for Density Ratio Learning}
\vspace{-0.5em}
Let $P$ and $Q$ be unknown probability measures \rnew{on a common sample space $\Omega$}. We focus on the case where $\Omega$ is a $d$-dimensional Euclidean space.
\rnew{We assume that $P$ and $Q$ admit densities $p$ and $q$, respectively, with respect to the Lebesgue measure $\mu$.}
Our goal is to estimate the unknown density ratio $p/q$ based on two i.i.d. samples.

Let $r$ denote an estimate of the density ratio. In the following discussion, we assume that $r$ is positive $\mu$-almost everywhere. 
To introduce an algorithm for optimizing $r$, we define a loss function called the balancing loss based on the following fact:
\nanew{If $r^*$ is equal to the true density ratio $p/q$,} then for any measurable subset $B \subset \Omega$, we have
\vspace{-1.5em}
\begin{align}
        \int_B (r^*)^{-1/2} p\, d\mu
    =
    \int_B (r^*)^{1/2} q\, d\mu
    = 
    \int_B \sqrt{pq}\,d\mu. \label{eq: balance in a subset}
\end{align}
\vspace{-3.5em}

\noindent
When $B=\Omega$, the whole sample space, this corresponds to the affinity between $p$ and $q$. The first equation implies that if we weight the two samples respectively with $(r^*)^{-1/2}$ and $(r^*)^{1/2}$,  
their weighted measures over any measurable set $B$ become equal.
In this sense, we say that the two samples are ``balanced''.
Motivated by this observation, we introduce the notation $w = r^{1/2}$ to denote the {\em balancing} function, which is intended to approximate the square root of the density ratio, $\sqrt{p/q}$.
We then consider the following {\em balancing} loss function:
\vspace{-1.5em}
\begin{align*}
    l(w)
    =
    \mathbb{E}_p [w^{-1}] + \mathbb{E}_q [w].
\end{align*}
\vspace{-3.5em}

\noindent
\nanew{
    Note that by the arithmetic-geometric mean inequality, the loss is bounded below by $2 \int \sqrt{pq} \, d \mu$, and equality holds if $w^{-1} p = w q$, or equivalently, $w = \sqrt{p/q}$.
    This simple justification is an advantage of selecting this functional form compared to other choices that can follow Eq.~\eqref{eq: balance in a subset}, such as
    $(\mathbb{E}_p[w^{-1}] - \mathbb{E}_q[w])^2$.
}

In practice, suppose we observe i.i.d.\ draws from $p$ and $q$: $x^{(0)}_1,\dots,x^{(0)}_{n_{0}} \sim p,\ x^{(1)}_1,\dots,x^{(1)}_{n_{1}} \sim\nobreak q$.

\Needspace{4\baselineskip}
Then the finite-sample version of the balancing loss is defined by
\vspace{-1em}
\[
    l_n(w)
    =
    \frac{1}{n_0}
    \sum^{n_0}_{i=1}
    w^{-1}(x^{(0)}_i)
    +
    \frac{1}{n_1}
    \sum^{n_1}_{i=1}
    w(x^{(1)}_i),
\]
\vspace{-3em}

\noindent
where $n = n_0 + n_1$ indicates the total number of observations.
\vspace{-1em}
\subsection{Connections to Existing Concepts in Statistical Learning}
\label{subsec: Connections to Existing Concepts in Statistical Learning}
\vspace{-0.5em}

While the design of the balancing loss may appear {\em ad hoc} at first, we next show that it can be naturally motivated from two well-established perspectives in machine learning. 

\vspace{-1em}
\subsubsection*{Connection 1: The ``Density-ratio Trick'' Based on Binary Classification}
\vspace{-0.5em}

A first connection is to the ``density-ratio trick'', which estimates the density ratio by inverting a binary classifier trained to distinguish the two samples \citep{sugiyama2012density, durkan2020contrastive, thomas2022likelihood}.
As we discuss below, the balancing loss arises naturally as a variant of the exponential loss, which is used for classification with AdaBoost \citep{friedman2000additive}, but only the balancing loss targets the density ratio directly. 

\rnew{To connect the exponential and balancing losses, suppose each observation is drawn from either sample with equal probability.}
That is, one first draws a random variable $y$ that takes 1 and -1 with a 50-50 chance. \rnew{Given $y$, draw $x$ from $P$ if $y=1$ and from $Q$ if $y=-1$.}
Then it is straightforward to show that $\mathbb{E}_p [w^{-1}] + \mathbb{E}_q [w] \propto \mathbb{E} [\exp(-y F)]$,
where $F = \log w$, and the expectation in the latter is taken with respect to the joint distribution of $x$ and $y$.
What we have on the right-hand side is exactly the exponential loss \citep{friedman2000additive, hastie2009elements}.
\rnew{However, the equivalence above no longer holds when the two sample sizes are unbalanced, because the exponential loss integrates $y$ with respect to its empirical marginal distribution rather than using a 50-50 split. 
Nevertheless, }because the population minimizer for AdaBoost is the log posterior odds between the two samples \citep{friedman2000additive}, theoretically we can estimate the density ratio via the density ratio trick \citep{sugiyama2012density}, namely,
\[
    \frac{p(x)}{q(x)}
    =
    \frac{\mathrm{Pr}(y=-1)}{\mathrm{Pr}(y=1)}
    \frac{\mathrm{Pr}(y=1 \mid x)}{\mathrm{Pr}(y=-1 \mid x)},
\]
\par\vspace{-0.5em}
\noindent where the prior ratio is replaced with the sample size ratio in practice.

\rnew{Although both approaches target the same density ratio in theory, they differ in practice for two reasons. 
First, boosting is regularized: it uses a small learning rate and is stopped before the loss is fully minimized, so the estimate remains close to its starting value.
The optimization of the exponential loss with AdaBoost typically starts from a posterior class probability of $1/2$, which corresponds to the density ratio $n_1/n_0$ after the prior-odds correction rather than to $p/q\equiv 1$. 
When the sample sizes are unbalanced, the density-ratio trick is thus shrunk toward a ratio that does not represent the absence of a difference. 
An additional experiment illustrating this bias is provided in Supplementary Materials~\ref{appendix: dr trick 1d}.
Second, an off-the-shelf classifier includes components designed for classification rather than for the density ratio, such as the impurity-based splitting rule for classification trees \citep{breiman1984classification} in the AdaBoost implementation \citep{culp2007ada}. These components can affect the accuracy of the estimate even when $n_0=n_1$, where both approaches shrink toward the same value.}

\nanew{
    \rnew{The balancing loss avoids both issues by construction. It is defined directly in terms of the density ratio, so its natural starting value $w\equiv 1$ represents no difference between the two samples and no post hoc correction is needed, and it leads to boosting algorithms whose components are tailored to the density ratio. Equivalently, its two group-specific terms are averaged separately, so the two groups receive equal total weight in the empirical objective.}
    Similar ideas of re-weighting the minority/majority groups are proposed in the context of classification for unbalanced data, as, for example, the cost-sensitive boosting \citep{sun2007cost}\rnew{, and with class weights inversely proportional to the sample sizes, the weighted exponential loss coincides with the balancing loss (see Supplementary Materials~\ref{appendix: loss relationship}).}
    To the best of our knowledge, the literature has not discussed either its application to density ratio estimation or, more importantly, its extension to Bayesian inference.
}

\vspace{-1em}

\subsubsection*{Connection~2: The Variational Form of the Squared Hellinger Distance}

\vspace{-0.5em}

\rnew{Estimating an $f$-divergence between $p$ and $q$ also motivates the balancing loss.} \rnew{Specifically, we show its exact equivalence to the variational form of the squared Hellinger distance.}

For a given convex function $f:[0,\infty)\to (-\infty,\infty]$ with $f(1)=0$ \rnew{and assuming that $P$ is absolutely continuous with respect to $Q$}, the general $f$-divergence is defined as    $D_f(P\|Q):=\int f(\frac{p}{q})\, d Q$. It is known \citep{renyi1961measures,csiszar1964informationstheoretische,ali1966general,nguyen2010estimating,polyanskiy2025information} that the dual form can be written as an optimization problem with respect to a functional class $\mathcal{F}$ mapping from $\Omega$ to $\mathbb{R}$,
$D_f(P\|Q) \geq \sup_{g\in\mathcal{F}}\int [g\, d P - f_c(g)\, d Q]$,
where $f_c$ is the conjugate dual function \nanew{associated with} $f$, defined as $f_c(v):=\sup_{u\in\R}\{uv-f(u)\}$. \nanew{The equality holds when the subdifferential $\partial f(
p/q
)$ contains an element in $\mathcal{F}$. The subdifferential of $f$ is defined as $\partial f(t):= \left\{z\in\mathbb{R}:f(s)\geq f(t)+z(s-t),\forall s\in \mathbb{R}\right\}$. See details in Lemma 1 of \citet{nguyen2010estimating}.} The dual objective is referred to as the variational form of the $f$-divergence. In the case of the squared Hellinger divergence, $f(u) = \frac{1}{2}(\sqrt{u}-1)^2$, and the  dual objective is $\sup_{g<\frac{1}{2}}\Big\{\mathbb{E}_{P}[g(X)] \;-\;\mathbb{E}_{Q}\!\Big[\frac{g(X)}{1-2g(X)}\Big]
    \Big\}$ with the optimum value $g^* = \frac{1}{2}-\frac{1}{2}\sqrt{\frac{q}{p}}$, based on a direct application of Lemma 1 in \cite{nguyen2010estimating}.  Now let $w = (1-2g)^{-1}$, the variational form for the squared Hellinger divergence becomes
\vspace{-1em}
\begin{equation}
\sup_{w>0} \left\{1-\frac{1}{2}\int w^{-1} p\, d\mu  - \frac{1}{2}\int w q \, d\mu\right\} \nonumber
\end{equation}
\vspace{-3em}

\noindent
which is clearly equivalent to minimizing the balancing loss. In other words, learning the density ratio by minimizing the balancing loss is essentially estimating the squared Hellinger distance using its variational form. Based on the known minimizer of the variational form, the minimizer of the balancing loss is $w^*=\sqrt{p/q}$. Hence, learning the density ratio can be achieved by minimizing the balancing loss.

\rnew{Beyond these two motivations, the symmetry of the balancing loss in $w$ vs $w^{-1}$ and $p$ vs $q$ will also facilitate generalized Bayesian inference on additive tree models.}

\vspace{-1em}
\subsection{Additive Tree Models and Boosting}
\vspace{-0.5em}

Given the close relationship between density ratios and posterior odds, \rnew{which can be estimated using supervised additive tree boosting}, \rnew{additive tree models are a natural candidate for approximating the density ratio}.
\rnew{We therefore fit} additive tree models to the density ratio under the balancing loss.
In this section, we begin with a non-Bayesian approach that aims to find a (functional) point estimate of the density ratio by optimizing the finite-sample version of the balancing loss. The resulting algorithms closely resemble supervised tree boosting.

Before introducing the algorithms, we describe the additive tree model. We consider a model for the log-balancing weight $\log w$ in the form
\vspace{-1.5em}
\begin{align}
\label{eq:additive_model}
\log w = \sum_{k=1}^{K} f_k
\end{align}
\vspace{-3.5em}

\noindent
where each base learner $f_k$ is a piecewise constant function defined on the leaf partition of the sample space given a tree $T_k$ with leaf node parameters $\beta_k = \{\beta_k(A)\}_{A \in T_k}$ and written as
\vspace{-1.5em}
\begin{align}
\label{eq:base_learner}
f_k = \sum_{A\in T_k} \beta_k(A) \1_A.
\end{align}
\vspace{-3.5em}

\noindent
Under this model, we are now presented with the following optimization problem:
\vspace{-1.5em}
\[
\min_{\{f_k\}} l_n(w) =  \min_{\{\beta_k,T_k\}} \frac{1}{n_0}
    \sum^{n_0}_{i=1}
    w^{-1}(x^{(0)}_i)
    +
    \frac{1}{n_1}
    \sum^{n_1}_{i=1}
    w(x^{(1)}_i)
\]
\vspace{-3.5em}

\rnew{We introduce two algorithms for fitting this model under the balancing loss.} \rnew{They directly parallel two approaches to fitting additive trees in supervised learning.} The first corresponds to a forward-stagewise (FS) algorithm based on greedy steepest descent on the space of single-tree-based piecewise-constant functions; the other corresponds to \citet{friedman2001greedy}'s gradient boosting (GB) algorithm. \rnew{In practice, these algorithms typically produce very similar estimates}, and so one could just adopt the simpler GB algorithm, but because the details of the FS algorithm shed more light on what constitutes effective function approximators for the density ratio under the balancing loss, we believe they are worthy of elaboration here.
\vspace{-1em}
\subsubsection*{Algorithm 1: A Forward-stagewise Algorithm}
\vspace{-0.5em}
The following proposition characterizes the optimal single tree fit in each iteration of a forward-stagewise algorithm, which takes $K$ steps of iterative fitting and in each step $k$ finds the optimal fit of $f_k$ that minimizes the balancing loss. \rnew{Its proof is provided in Supplementary Materials~\ref{appendix: proofs}.}

\begin{prop}
    \label{prop: optimal trees for adaboost (for manucscript)}
    Suppose $w_{k-1}$ is the estimate of \nanew{$\sqrt{r^*} = \sqrt{p/q}$} after $(k-1)$ steps of fitting. Then in the $k$th step, the loss is minimized by $\log w_{k}=\log w_{k-1}+f_k$ when the following conditions are satisfied by $f_k$:
    \begin{enumerate}
    \item Let $P_k |_{T_k}$ and $Q_k |_{T_k}$ be discrete distributions with their probability masses indexed by $A \in T_k$ such that
    \vspace{-2em}
    \[
        P_k |_{T_k} (A)
        \propto 
        \int_A w_{k-1}^{-1} p\, d\mu,\ 
        Q_k |_{T_k} (A)
        \propto 
        \int_A w_{k-1} q\, d\mu.
    \]
    \vspace{-4em}

    \noindent
    Then $T_k$ is a partition structure that maximizes the Hellinger distance
    \vspace{-1.5em}
    \[
    H(P_k |_{T_k}, Q_k |_{T_k})
    =
    \frac{1}{\sqrt{2}}
    \sqrt{
        \sum_{A \in T_k}
        \left(
            \rnew{\sqrt{P_k |_{T_k}(A)} - \sqrt{Q_k |_{T_k}(A)}}
        \right)^2.
    }
    \]
    \vspace{-3.5em}
    \item For each $A \in T_k$, $\exp(\beta_k(A)) = \sqrt{\int_A p w^{-1}_{k-1}\, d\mu }/\sqrt{\int_A q w_{k-1}\, d\mu}$.
    \end{enumerate}    
\vspace{-2em}
\end{prop}
\vspace{1em}

\noindent
{\it Remark 1}: If the balancing weight is correctly estimated, it achieves the symmetry between the two expectations in the loss in that $\mathbb{E}_p [w^{-1}] = \mathbb{E}_q [w]$, and we can impose this condition as a restriction or regularization on the estimator so that it does not focus on minimizing only one expectation out of the two. 
As justification, Proposition~\ref{prop: optimal trees for adaboost (for manucscript)} implies that the balancing loss can be improved by multiplying the current estimate $w_k$ by a constant ($c$, say) such that $\mathbb{E}_p [(c w_k)^{-1}] = \mathbb{E}_q [c w_k]$ (\rnew{see Supplementary Materials~\ref{appendix: proofs} for details}).
Based on this discussion, we introduce this correction step in the forward-stagewise algorithm.
\vspace{0.5em}

\noindent {\it Remark 2}:  Given the connection between the balancing loss and the variational form of the Hellinger divergence, the ``optimal'' single-tree in each iteration naturally corresponds to the partition that maximizes the discretized approximation of the Hellinger divergence.

As in tree-based predictive models, searching for the exact optimal tree in each step is often not computationally tractable for multivariate data. Instead one can adopt a greedy, coarse-to-fine algorithm in which we iteratively choose a single cut point at each split of the tree $T_k$ that achieves the steepest ascent in the Hellinger distance or the steepest decrease in the corresponding affinity. Specifically, when we consider splitting the current node $A$, we can search for the splitting rule such that for the child nodes $A_l$ and $A_r$ \nanew{the following Hellinger affinity is minimized, or, equivalently, the Hellinger distance is maximized}:
\vspace{-1em}
\[ 
    \sqrt{
    P_k |_{T_k}(A_l)
    Q_k |_{T_k}(A_l)
    }
    +
    \sqrt{
    P_k |_{T_k}(A_r)
    Q_k |_{T_k}(A_r)
    }.
\]

\vspace{-2em}

\vspace{-0.5em}
\subsubsection*{Algorithm 2: A Gradient Boosting Algorithm}
\vspace{-0.5em}
We can also introduce a gradient boosting algorithm \citep{friedman2001greedy}, noting that the negative gradients of the empirical loss with respect to the additive estimand $F$ are as follows:
\vspace{-1em}
\[
    -\frac{\partial l_n(w)}{\partial F(x_i^{(0)})}=\frac{1}{n_{0}}\exp(-F(x_i^{(0)})) = \frac{1}{n_0} w^{-1}(x_i^{(0)}),\mkern-1mu
    -\frac{\partial l_n(w)}{\partial F(x_i^{(1)})}=\rnew{-\frac{1}{n_1}\exp(F(x_i^{(1)}))} = -\frac{1}{n_1}w(x_i^{(1)}),
\]
\vspace{-3em}

\noindent
which are the pseudo-residuals. One can then follow the procedure of \citet{friedman2001greedy}. Specifically, in each iteration of the algorithm we fit a regression tree to the pseudo-residuals by minimizing the squared error loss, that is, minimizing the variances of the gradients in the leaf nodes. We again adopt a coarse-to-fine greedy search and for each node $A$, we select the splitting rule with child nodes $A_l$ and $A_r$ that minimizes the variances.
After constructing the tree, we obtain the optimal values of the node parameters given by Proposition~\ref{prop: optimal trees for adaboost (for manucscript)}.
\vspace{-1.0em}
\subsubsection*{Regularization Strategies}
\vspace{-0.5em}
\rnew{As in supervised learning, both boosting algorithms require regularizing each base learner to extract only a small amount of density-ratio structure.}
\rnew{We regularize each learner's contribution using a learning rate} $\nu \in (0,1]$.
If the current estimate is $w_{k-1}$ and the optimal learner obtained in the $k$th step is $f_k$, we can combine them as follows: $\log w_{k} = \log w_{k-1} + \nu f_k$.

It is known that smaller values of the learning rate $\nu$ lead to better generalization performance in both supervised learning \citep{friedman2001greedy, hastie2009elements} and unsupervised learning \citep{awaya2024unsupervised}.
Accordingly, in our implementation, we follow them by setting $\nu$ to, for example, $0.01$. 

\rnew{A complementary regularization strategy limits tree size to avoid overly complicated partitions; for example, we set the maximum tree resolution (depth) to 4.}
Finally, the number of trees in the ensemble also needs to be carefully selected. We find that choosing this number via cross-validation yields satisfactory performance. 
\vspace{-1em}
\section{Generalized Bayesian Inference on Additive Trees}
\vspace{-0.5em}

\nanew{
Tree boosting only offers point estimates for the density ratio function, without providing any associated uncertainty. To achieve uncertainty quantification, next we introduce a generalized Bayesian strategy \citep{bissiri2016general} for learning the density ratio. 
}
\vspace{-1.5em}
\subsection{Balancing Loss as a Pseudo-likelihood}
\vspace{-0.5em}
\nanew{Following the framework introduced in \citep{bissiri2016general}}, we define a pseudo-likelihood using the finite-sample loss $l_n(w)$, that is,
\nanew{
\vspace{-1.5em}
\[
   L_{n,\tau} (w)
    =
    \tau^{2 n_\mathrm{min}}
    \exp (- n_{\mathrm{min}} \tau l_n(w))
\]
\par\vspace{-1.5em}
\noindent where $n_{\min}$ is the sample size of the smaller group $\min(n_0, n_1)$, and
$\tau>0$ is a ``temperature'' parameter.
The term $\tau^{2 n_\mathrm{min}}$ is added to provide a clearer interpretation of the pseudo-likelihood as a weighted likelihood under exponential distributions, as follows:
\vspace{-0.5em}
\begin{align}
L_{n,\tau}(w)
=
\prod^{n_0}_{i=1}
\left[
\tau
\exp
\left(
    -
    \tau
    w^{-1}
    (x^{(0)}_{i})
\right)
\right]^{\zeta_0}
\times 
\prod^{n_1}_{i=1}
\left[
\tau
\exp
\left(
    -
    \tau
    w
    (x^{(1)}_{i})
\right)
\right]^{\zeta_1},
\label{eq: weighted likelihood expression}
\end{align}
\par\vspace{-1.5em}
\noindent where $\zeta_0 = n_{\mathrm{min}}/n_0$ and $\zeta_1 = n_{\mathrm{min}}/n_1$ are the weights taking values in $(0,1]$.
\rnew{If $n_0 < n_1$, for example, the second group's observations are down-weighted to balance information from the two groups.}
The effective sample size is controlled not by the total sample size but by $n_\mathrm{min}$, \rnew{reflecting the dependence of information about the two-sample difference on the smaller group size.}
}

\nanew{
    \rnew{Under this formulation}, the temperature $\tau$ can also be interpreted by a one-dimensional summary of the two-sample difference.
    To see this, we note that the likelihood $L_{n,\tau}(w)$ is maximized with respect to $\tau$ when $\tau^{-1} = \frac{1}{2} l_n (w)$.
    As $w$ is the balancing function corresponding to 
    $\sqrt{p/q}$, what we estimate with $\tau^{-1}$ is $\frac{1}{2} l_n (\sqrt{p/q}) \approx \int \sqrt{pq}\, d\mu$,
    which is the Bhattacharyya coefficient \citep{bhattacharyya1946measure}.
}

Given the prior distribution of $w$, $\pi_0(w)$, the posterior distribution of $w$ is defined as $\pi(w) \propto \pi_0(w) L_{n,\tau} (w)$,
\nanew{
where $\pi_0(w)$ is the prior of the balancing function $w$, specified next.
}

\vspace{-1.5em}

\Needspace{4\baselineskip}\subsection{Conjugate Prior Specification}
\label{sec: The prior distribution of the density ratio and Gibbs sampler}
\vspace{-0.5em}

\rnew{We describe priors on $w$ for efficient MCMC and a data-driven approach to setting $\tau$.} Recalling our model introduced in Eq.~\eqref{eq:additive_model}, we describe a Gibbs sampling strategy that iteratively samples from the full conditional of each basis function $f_k$, given all the others. We update the leaf node parameters $\beta_k(A)$ and the partition $T_k$ defining $f_k$ in Eq.~\eqref{eq:base_learner}.

\vspace{-1em}
\subsubsection{\rnew{Updating the Leaf Node Parameters}}
\vspace{-0.5em}
To derive the full conditional for $\beta_k(A)$, we define the balancing weight with $f_k$ subtracted: $\log w_{-k} := \log w - f_k$.

\Needspace{4\baselineskip}
The portion of the likelihood $L_{n,\tau} (w)$ that depends on $\beta_k(A)$ is given by
\vspace{-1.5em}
\[
    L_{n,\tau} (w)
    \propto
    \exp
    \left(
        -   
        \nanew{\zeta_0}
        \tau
        \gamma_k(A)^{-1}
        \sum_{x^{(0)}_i \in A} w_{-k}^{-1} (x^{(0)}_i)
        -
      \nanew{\zeta_1}
        \tau
        \gamma_k(A)
        \sum_{x^{(1)}_i \in A} w_{-k} (x^{(1)}_i)       
    \right),
\]
\vspace{-2.5em}

\noindent
where $\gamma_k(A) = \exp(\beta_k(A))$. Noting that this pseudo-likelihood resembles an exponential family kernel for $\gamma_k(A)$, \nanew{we adopt an inverse-Gaussian leaf prior \citep{murray2021log}} for $\gamma_k(A)$, which is conjugate to this kernel. Specifically, the prior $\mathrm{InverseGauss}(\mu_k(A), \lambda_k(A))$  on $\gamma_k(A)$ is:
\vspace{-1em}
\begin{align*}
    p(\gamma_k(A) = z \mid \mu_k(A), \lambda_k(A))
    &=
    \sqrt{
        \frac{\lambda_k(A)}{2 \pi z^3}
    }
    \exp
    \left(
        - \frac{\lambda_k(A) (z - \mu_k(A))^2}{2 \mu_k(A)^2 z }
    \right)\\
    &\propto 
    \sqrt{
    \frac{\lambda_k(A)}{2 \pi z^3}
    }
    \exp
    \left(
        -\frac{\lambda_k(A)}{2\mu_k(A)^2}z
        -\frac{\lambda_k(A)}{2z}
    \right).
\end{align*}
\vspace{-3em}

\noindent
It is straightforward to see that the conditional posterior is analytically obtained in the form of $\mathrm{InverseGauss}(\mu'_k(A), \lambda'_k(A))$, where
\vspace{-1em}
\begin{align*}
    \mu'_k(A)
    =
    \sqrt{
    \frac{\lambda_k(A) + 2 \nanew{\zeta_0} \tau \sum_{x^{(0)}_i \in A} w_{-k}^{-1} (x^{(0)}_i)}
    {\frac{\lambda_k(A)}{\mu_k(A)^2} + 2 \nanew{\zeta_1} \tau \sum_{x^{(1)}_i \in A} w_{-k} (x^{(1)}_i) } 
    },\
    \lambda'_k(A) =
    \lambda_k(A) 
    +
    2
    \nanew{\zeta_0}
    \tau
    \sum_{x^{(0)}_i \in A} w_{-k}^{-1} (x^{(0)}_i).
\end{align*}
\vspace{-3em}

\noindent
We set $\mu_k(A)$ to $1$ so that the induced marginal prior on $\log w$ is around 0 (more discussions are to be given at the end of this section).
The precision parameter $\lambda_k(A)$ induces shrinkage toward the prior mean~1, as seen in the expression of the posterior mean $\mu'_k(A)$.
We set $\lambda_k(A)$ according to the number of trees $K$, namely, $
    \lambda_k(A)
    =
    \lambda_0 K,$
where $\lambda_0$ is a hyperparameter.  
The prior value of $\lambda_k(A)$ is linear in $K$ because the results in \cite{whitmore1978normalizing} imply that the variance of $f_k (x)$ for each point $x$ ($x \in \Omega$) is $O(K^{-1})$ under our model specification (\rnew{see Supplementary Materials~\ref{appendix: prior derivation} for more details}). This choice of $\lambda_k(A)$ is also analogous to the amount of shrinkage recommended for the BART model \citep{chipman2010bart}.
We set $\lambda_0$ to~5 as a default value in all of our numerical experiments.

Another consideration is that the inverse Gaussian distribution has an asymmetric shape, so it can also introduce skewness into the prior of the log-balancing weight $\log w$.
To avoid this issue, we apply the inverse Gaussian prior to $\gamma_k(A)$ and $\gamma_k(A)^{-1}$ alternately, that is,
$\gamma_k(A) \sim \mathrm{InverseGauss} (1,\lambda_0 K)\ (k \text{: odd}),\ \gamma_k(A)^{-1} \sim \mathrm{InverseGauss} (1, \lambda_0 K)\ (k \text{: even})$.
This ensures that the marginal prior of $\log w$ is exactly centered at $0$ whenever the total number of trees $K$ is even.
\vspace{-1em}
\subsubsection{Tree Priors and Posterior Updates on the Trees}
\vspace{-0.5
em}

We adopt the standard Bayesian CART prior \citep{chipman1998bayesian} on the tree $T_k$ as in BART \citep{chipman2010bart} \rnew{using a recursive partition process}. \rnew{Starting with the full sample space as the root, each node $A$ at depth} $\mathrm{depth}(A)$ \rnew{(0 for the root)} is further split with probability $a_T (1 + \mathrm{depth}(A))^{-b_T},\ a_T \in (0,1), b_T \in [0,\infty)$.

\rnew{We use the standard default hyperparameters} $a_T = 0.95$ and $b_T=2$ \citep{chipman2010bart}.
\rnew{To split a node, we select one of the $d$ dimensions with equal probability and a cut point uniformly over its range.}
Following the standard recipe for BART \citep{chipman2010bart, kapelner2016bartmachine, sparapani2021nonparametric}, we update the tree $T_k$ by using Metropolis-Hastings moves. \rnew{The details of the tree updating moves are provided in Supplementary Materials~\ref{appendix: Details of the MCMC algorithm}.}
\vspace{-1em}
\subsection{Tuning the Temperature $\tau$}
\vspace{-0.5em}
In generalized Bayesian inference, the temperature parameter $\tau$ controls the strength of the loss-based likelihood for the balancing weight function $w$ relative to the prior and significantly influences the resulting posteriors, including posterior variances and credible intervals.
\rnew{Methods for empirical temperature selection include} information matching \citep{holmes2017assigning}, calibrating frequentist coverage rates of credible intervals \citep{syring2019calibrating}, and the SafeBayes algorithm \citep{grunwald2017inconsistency}.
\rnew{Adapting these methods is difficult here because we learn an unknown, nonparametrically specified function in an unsupervised setting with relatively costly MCMC.}

Instead, we adopt a hierarchical Bayesian approach by placing a prior on the temperature $\tau$, following \cite{bissiri2016general, rigon2023generalized}. \nanew{Motivated by the weighted likelihood expression in Eq.~\eqref{eq: weighted likelihood expression},
we adopt a conjugate prior $\pi_0(\tau)$, namely  $\mathrm{Gamma} (a_{0,\tau}, b_{0,\tau})$, so the full conditional of $\tau$ is $\mathrm{Gamma} (a_{1,\tau}, b_{1,\tau})$, where $a_{1,\tau} = a_{0,\tau} + 2 n_{\mathrm{min}},\ b_{1,\tau}= b_{0,\tau} + n_{\mathrm{min}} l_n(w).$
}

\begin{figure}[htb]
  \centering
  \includegraphics[width=\linewidth]{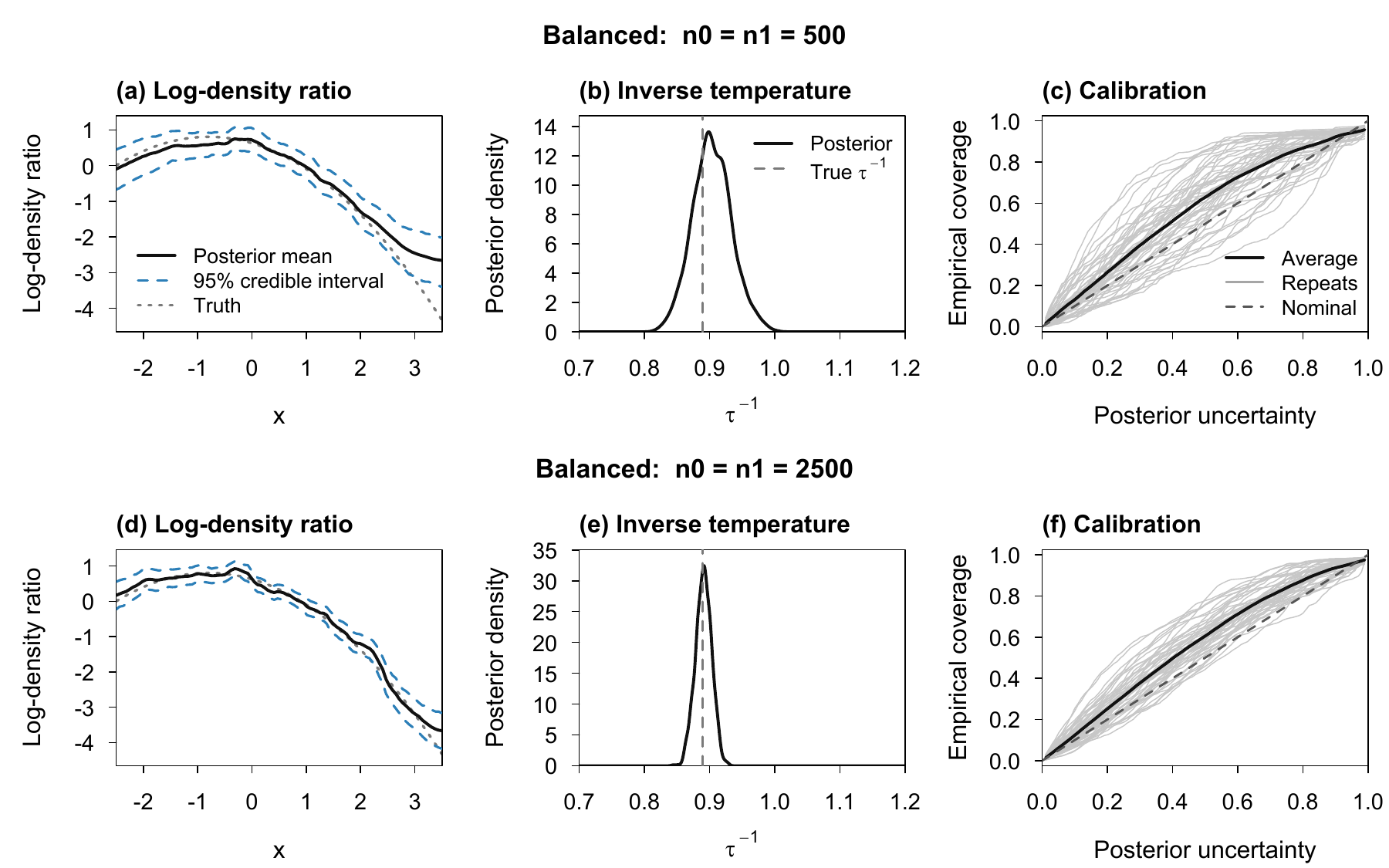}
  \caption{
  \nanew{
  \rnew{An evaluation of posterior distributions under the 1D scenarios with balanced sample sizes, $(n_0,n_1)=(500,500)$ and $(2500,2500)$.}
  \rnew{Left: true log-density ratios} (gray, dotted) \rnew{and posteriors} (mean: black, solid, 95\% interval: blue, dashed) evaluated pointwise.
  \rnew{Middle: posterior distributions of inverse temperature} $\tau^{-1}$ and the true Bhattacharyya coefficient.
  \rnew{Right: calibration plots comparing} the posterior uncertainty and the nominal coverage rate (namely, the ratio of the true log-density ratios included in the pointwise credible intervals) based on 50 simulations and their averages.
  }
  }
  \label{fig: 1D illustrations}
\end{figure}

\vspace{-1.0em}
\nanew{
\subsection{Empirical Validation of Uncertainty Calibration}
\vspace{-0.5em}
\label{sec: uncertainty calibration}
\rnew{We assess the calibration of generalized Bayesian uncertainty using two one-dimensional Gaussian distributions, $N(0, 1)$ and $N(1, 2.25)$.}

\rnew{The posterior distributions under two balanced sample sizes are provided in Figure~\ref{fig: 1D illustrations}, where the posterior means and the credible intervals are evaluated pointwise. The corresponding results under two unbalanced sample sizes are provided in Supplementary Figure~\ref{fig: 1D illustrations unbalanced}.}
\rnew{The posterior distributions are centered around the true log-density ratio where there are sufficient observations} (i.e., the region between -2 and 2).
Moreover, the inverse temperature $\tau^{-1}$, with which we estimate the Bhattacharyya coefficient, is distributed around the true value.
We note that more information about the two-sample difference is provided under the balanced sample size as is represented by the weighted likelihood expression, and it is reflected by the different widths of the credible intervals in the posterior distributions.
Figure~\ref{fig: 1D illustrations} also shows the calibration of the credible intervals obtained by our Bayesian model in terms of nominal coverage rates based on 50 simulated data sets.
\rnew{Posterior uncertainty and nominal coverage tend to agree more at larger sample sizes}, as in the case of $(n_0, n_1) = (2500, 2500)$.
The cases $(500,500)$ and $(500,4500)$ are similar, which is consistent with the loss-based likelihood being driven by the smaller group size.
\rnew{The corresponding calibration results for the two-dimensional scenarios are provided in Supplementary Figure~\ref{figure: coverage rate 2D}.}
}

\vspace{-1.5em}
\section{Numerical Experiments}
\label{sec: numerical experiments}
\vspace{-0.5em}
\rnew{We use simulations to evaluate both proposed methods: tree boosting and Bayesian additive trees.}
We begin with scenarios involving bivariate distributions,  
and then consider higher-dimensional (20-dimensional) cases, providing a comparison with several existing approaches. 
\rnew{The first comparator is the} ``density-ratio trick'' by inverting a binary classifier. \rnew{Its qualitative differences from our boosting algorithms are discussed in Section~\ref{subsec: Connections to Existing Concepts in Statistical Learning}.} In particular, we adopt AdaBoost for binary classification, as it uses the same class of additive tree ensembles to classify the two classes using essentially the exponential loss \citep{friedman2000additive}.
\rnew{For comparability, the two implementations adopt the same tuning parameter settings, including the learning rate, the tree depth, and the fraction of observations randomly subsampled to construct each tree (0.5), and their implementation details are summarized in Supplementary Materials~\ref{appendix: details of experiments}.}
To apply the density-ratio trick, we first estimate the classification probabilities and then transform them into the density ratios by multiplying the sample size ratios. \rnew{We use the} \texttt{ada} function provided in the R package \texttt{ada} \citep{ada_package}.
\nanew{In addition to the density-ratio trick, we evaluate the calibration method proposed in \cite{cranmer2015approximating} using AdaBoost as a discriminator.}
\rnew{For the AdaBoost-based estimators, we select the number of trees by five-fold cross-validation based on the exponential loss.}
We also compare our methods with two popular kernel-based density ratio estimators, namely KLIEP \citep{sugiyama2008direct} and uLSIF \citep{kanamori2009least}, implemented with the R package \texttt{densratio} \citep{densratio_package}.
For the proposed methods, we assess both types of boosting, the FS and the GB algorithms, along with the Bayesian additive tree model.
\rnew{Implementation details for all estimators are provided in Supplementary Materials~\ref{appendix: details of experiments}.}

For each scenario, we evaluate the performance under two sample size settings: (i) $n_0 = n_1 = 5000$ (the ``balanced'' setting) and (ii) $n_0 = 9000$, $n_1 = 1000$ (the ``unbalanced'' setting). 
As a measure of accuracy, we use the following equally weighted mean squared error of the log-density ratio for the true density ratio $r^*$ and the estimated ratio $\hat{r}$:
\vspace{-1em}
\[
    \frac{1}{2 n_0}
    \sum^{n_0}_{i=1}
        \left(
        \log r^*(x_i^{(0)})
        - 
        \log \hat{r}(x_i^{(0)})
    \right)^2
    +
    \frac{1}{2 n_1}
    \sum^{n_1}_{i=1}
        \left(
        \log r^*(x_i^{(1)})
        - 
        \log \hat{r}(x_i^{(1)})
    \right)^2,
\]
\vspace{-3em}

\noindent
and for each setting we calculate \nanew{the average error defined above} based on 50 simulated datasets.

\vspace{-1em}
\subsection{Two-dimensional Simulations}
\label{sec: 2d scenarios}
\vspace{-0.5em}

\begin{figure}[htb]
\centering
\begin{tabular}{c}
    \includegraphics[height=5cm]{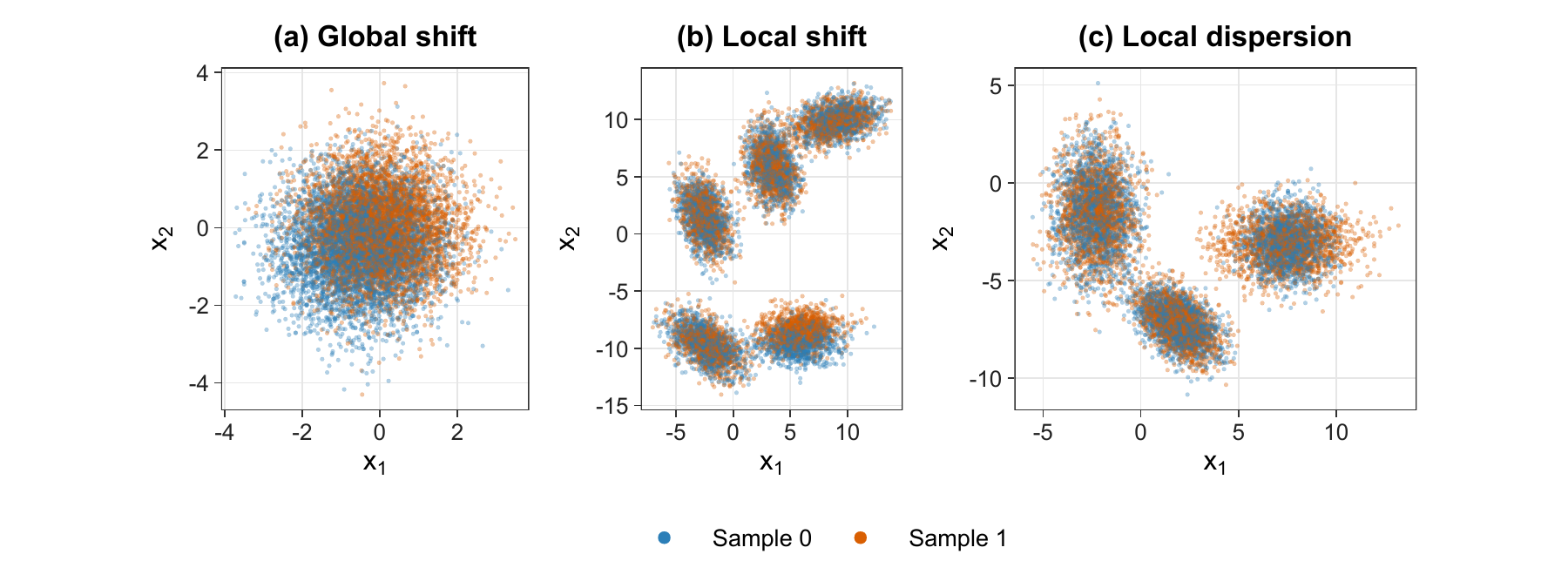}
\end{tabular}
\caption{A representative sample of simulations generated from the three scenarios used for the two-dimensional experiments. 
The sample sizes, $n_0$ and $n_1$, are both set to 5000.
}
\label{figure: 2d scenarios}
\end{figure}

We simulate data sets using the following three scenarios which represent different types of distributional differences between $X_0$ and $X_1$ visualized in Figure~\ref{figure: 2d scenarios}. 
The first scenario represents a global-scale difference between the two samples, whereas the second and third scenarios involve more localized discrepancies.
\rnew{The details are provided in Supplementary Materials~\ref{appendix: details of experiments}.}

\rnew{Table~\ref{table: MSE(2D)} compares the MSEs across the two-dimensional scenarios, and we can see that the lowest MSE in each setting is obtained by one of the proposed methods. 
For DRT, the increase in MSE from the balanced to the unbalanced setting is larger than those for GB and FS in all three scenarios. 
CDC does not improve upon DRT in these experiments and also produces estimates with infinite values for some data sets; the MSEs based on the finite estimates are shown in the table, and details are provided in Supplementary Materials~\ref{appendix: additional figures}.}

\begin{table}[htb]
\centering
    \caption{A comparison of the MSEs and their standard errors in the two-dimensional scenarios.
    }\label{table: MSE(2D)}
\begin{tabular}{lcccccc}\toprule
    &\multicolumn{2}{c}{\textbf{Global Shift}}&\multicolumn{2}{c}{\textbf{Local Shift}}&\multicolumn{2}{c}{\textbf{Local Dispersion}}
    \\\cmidrule(r){2-3}\cmidrule(r){4-5}\cmidrule(r){6-7} 
     $n_0 / (n_0 + n_1)$ &0.5&0.9&0.5&0.9&0.5&0.9 \\\midrule
GB &\rnew{0.025} &\rnew{0.049} &\rnew{0.036} &\rnew{0.068} &\rnew{0.105} &\rnew{0.130} \\
&\rnew{(0.001)} &\rnew{(0.001)} &\rnew{(0.001)} &\rnew{(0.002)} &\rnew{(0.002)} &\rnew{(0.005)} \\
FS &\rnew{0.027} &\rnew{0.052} &\rnew{0.034} &\rnew{0.069} &\rnew{0.103} &\rnew{0.132} \\
&\rnew{(0.001)} &\rnew{(0.001)} &\rnew{(0.001)} &\rnew{(0.002)} &\rnew{(0.002)} &\rnew{(0.005)} \\
BAT &\rnew{0.010} &\rnew{0.019} &\rnew{0.046} &\rnew{0.139} &\rnew{0.112} &\rnew{0.192} \\
&\rnew{(0.000)} &\rnew{(0.001)} &\rnew{(0.001)} &\rnew{(0.002)} &\rnew{(0.002)} &\rnew{(0.006)} \\
DRT (AdaBoost) &\rnew{0.043} &\rnew{0.107} &\rnew{0.047} &\rnew{0.118} &\rnew{0.121} &\rnew{0.180} \\
&\rnew{(0.001)} &\rnew{(0.003)} &\rnew{(0.001)} &\rnew{(0.002)} &\rnew{(0.002)} &\rnew{(0.004)} \\
CDC (AdaBoost) &\rnew{0.207} &\rnew{0.633} &\rnew{1.161} &\rnew{1.522} &\rnew{1.296} &\rnew{2.112} \\
&\rnew{(0.009)} &\rnew{(0.026)} &\rnew{(0.139)} &\rnew{(0.165)} &\rnew{(0.192)} &\rnew{(0.190)} \\
KLIEP &\rnew{0.151} &\rnew{0.158} &\rnew{0.307} &\rnew{0.312} &\rnew{0.303} &\rnew{0.308} \\
&\rnew{(0.005)} &\rnew{(0.006)} &\rnew{(0.004)} &\rnew{(0.005)} &\rnew{(0.004)} &\rnew{(0.007)} \\
uLSIF &\rnew{1.121} &\rnew{0.482} &\rnew{0.388} &\rnew{0.389} &\rnew{0.149} &\rnew{0.198} \\
&\rnew{(0.168)} &\rnew{(0.104)} &\rnew{(0.017)} &\rnew{(0.021)} &\rnew{(0.007)} &\rnew{(0.010)} \\
    \bottomrule
\end{tabular}
\begin{tablenotes}
\small
\item Notes: The methods are GB (gradient boosting), FS (forward stagewise), BAT (Bayesian additive trees), DRT \nanew{and CDC} (density-ratio trick and \nanew{calibrated discriminative classifier} based on AdaBoost), and the two kernel-based methods KLIEP and uLSIF.
\end{tablenotes}
\end{table} 

\begin{figure}[!tbp]
\centering
\begin{tabular}{c}
    \includegraphics[width=13.5cm]{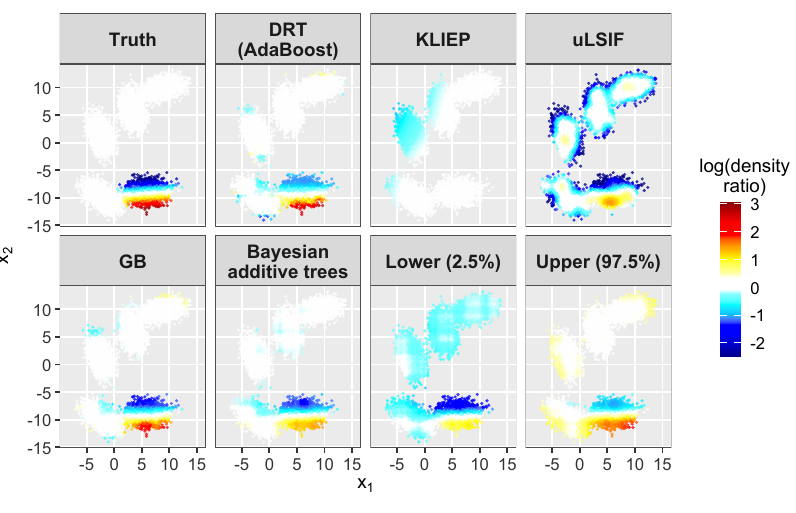}
\end{tabular}
\caption{The estimated log-density ratios obtained from the algorithms considered in the two-dimensional examples for the Local Shift scenario ($n_0 = n_1 = 5000$).
For Bayesian additive trees, the pointwise 2.5\% quantiles and 97.5\% quantiles of the log-density ratios are also displayed. The difference between FS and GB is minimal, so we show only the results for GB.}
\label{figure: 2d visualization of the estimated ratios}
\end{figure}

\rnew{Figure~\ref{figure: 2d visualization of the estimated ratios} compares the estimated log-density ratios in the Local Shift scenario with balanced sample sizes, $n_0=n_1=5000$. 
\rnew{The tree-based estimators (DRT and the proposed methods) identify the shifted mixture component as the primary region with nonzero log-density ratios.
Among the kernel-based methods, KLIEP does not capture this region, whereas uLSIF captures it but also produces sizable nonzero estimates where the true log-density ratio is close to zero.}
For Bayesian additive trees, \rnew{the lower and upper posterior quantiles also identify pointwise credible intervals excluding zero}: yellow and red points in the lower-quantile panel correspond to intervals entirely above zero, while blue points in the upper-quantile panel correspond to intervals entirely below zero. \rnew{Visualizations for the other scenarios and sample sizes appear in} Supplementary Materials~\ref{appendix: additional figures}.}

\vspace{-1em}
\Needspace{4\baselineskip}\subsection{Higher-dimensional Settings}
\label{sec: higher-dimensional settings}
\vspace{-0.5em}

\rnew{We evaluate the density ratio estimation algorithms in higher dimensions.}
\rnew{We generate 20-dimensional observations from a latent factor model.} First, we simulate latent variables $Z$ from 4-dimensional distributions, introducing the difference between the two samples in this latent space, and then project them onto the 20-dimensional space as
$\Lambda Z + \epsilon_i,\ \epsilon_i \sim N(\mathbf{0}, (0.1)^2 I_{20})$,
where $\Lambda$ is a randomly generated orthonormal $20 \times 4$ matrix satisfying $\Lambda' \Lambda = I_4$.
The distributions of the latent variables for the two groups, denoted by $Z_0$ and $Z_1$, are defined under \rnew{the following three scenarios.}
\vspace{-0.5em}
\begin{enumerate}[topsep=3pt,partopsep=0pt,parsep=0pt,itemsep=2pt]
    \item \rnew{``Global Shift'': $
        Z_0 \sim N(-\delta/2, I_4),\
        Z_1 \sim N(\delta/2, I_4)$,
    where $\delta=(1,0,0,0)^T$.}
    \item ``Location Shift'': $ Z_0
        \sim
        \frac{4}{5}
        N(\mu_1, \Sigma_1)
        +
        \frac{1}{5}
        N(\mu_2, \Sigma_2),\
        Z_1
        \sim
        \frac{4}{5}
        N(\mu_1 + \delta, \Sigma_1)
        +
        \frac{1}{5}
        N(\mu_2, \Sigma_2)$, where $\delta = (1,0,0,0)^T$.
    \item ``Dispersion Difference'': $ Z_0
        \sim \frac{4}{5}
        N(\mu_1, \Sigma_1)
        +
        \frac{1}{5}
        N(\mu_2, \Sigma_2),\
        Z_1
        \sim \frac{4}{5}
        \rnew{N(\mu_1, \Delta \Sigma_1 \Delta^{\mathsf T})}
        +
        \frac{1}{5}
        N(\mu_2, \Sigma_2)$, where $\Delta = \mathrm{diag}(0.5, 1,1,1)$. \rnew{Since $\Sigma_1=I_4$, the first latent coordinate in this component has variance $0.25$.}
\end{enumerate}
\rnew{The parameters $\mu_k$ and $\Sigma_k$ ($k=1,2$) used in the Location Shift and Dispersion Difference scenarios are provided in Supplementary Materials~\ref{appendix: details of experiments}.}
The orthonormal matrix $\Lambda$ is generated using the \texttt{randortho} function in the R package \texttt{pracma} \citep{pracma_package}.

\rnew{An illustrative comparison of the true and estimated log-density ratios in the latent space for the representative scenario (Location Shift) is provided in Supplementary Figure~\ref{fig:supp_20d_location_density_ratio}.}
The latent factor model for the true densities represents a challenging case for density ratio estimation, particularly for tree-based methods that use axis-aligned partitions along the original dimensions of the sample space.
\rnew{Table~\ref{table: MSE(multi)} shows a particularly clear contrast between GB and DRT, two boosting algorithms using additive trees with the same tuning parameter settings but different loss functions.
GB has lower MSE than DRT in all six settings, and the gap widens under sample-size imbalance: the difference is about $0.01$ in the three balanced scenarios but between $0.05$ and $0.08$ in the three unbalanced scenarios.
This pattern is consistent with the two differences discussed in Section~\ref{subsec: Connections to Existing Concepts in Statistical Learning}: the classification-oriented components of the AdaBoost implementation can affect the estimate for any sample sizes, whereas the shrinkage toward the prior-odds-corrected ratio $n_1/n_0$ arises only under sample-size imbalance.}

\begin{figure}[!tbp]
    \centering
    \includegraphics[width=13.5cm]{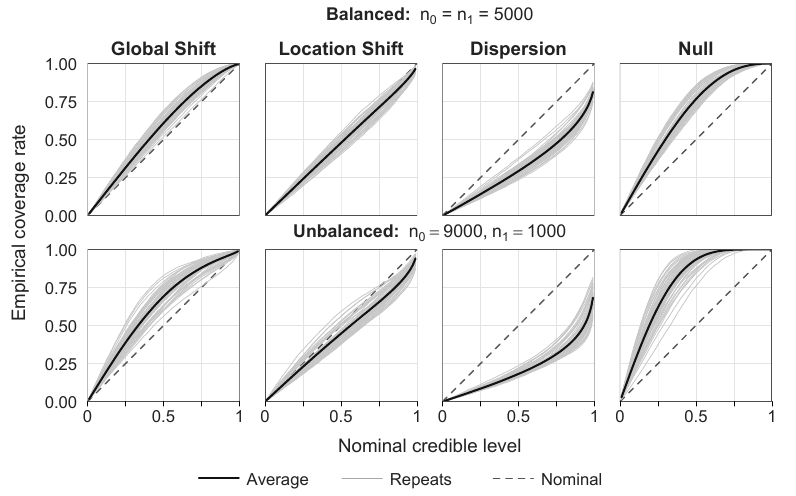}
    \caption{\rnew{
    Calibration of the pointwise credible intervals in the four 20-dimensional scenarios.
    \rnew{Columns show Global Shift, Location Shift, Dispersion, and Null; rows show balanced sample sizes} $(n_0,n_1)=(5000,5000)$ \rnew{and unbalanced sample sizes} $(n_0,n_1)=(9000,1000)$.
    For each nominal credible level, the empirical coverage rate is the proportion of evaluated observations whose pointwise credible interval contains the true log-density ratio.
    \rnew{Gray curves show the 50 simulated datasets, the black curve their average, and the dashed line equality between nominal credible level and empirical coverage.}
    }}
    \label{fig:20d-calibration}
\end{figure}

\rnew{
To evaluate the proposed BART method, Figure~\ref{fig:20d-calibration} examines the calibration of the pointwise credible intervals under the three scenarios and an additional scenario labeled ``Null,'' in which $Z_0$ and $Z_1$ both follow $N(0,I_4)$ under the same projection and noise model.
The Global Shift and Null scenarios exhibit conservative coverage, whereas the Location Shift scenario is substantially closer to the nominal coverage levels.
Undercoverage is \rnew{most evident} in the Dispersion scenario, and this is consistent with the greater difficulty of representing a dispersion difference introduced in the latent space and projected into the observation space using axis-aligned tree partitions, especially when one of the two samples is relatively small.}

\begin{figure}[!tbp]
    \centering
    \includegraphics[width=13.5cm]{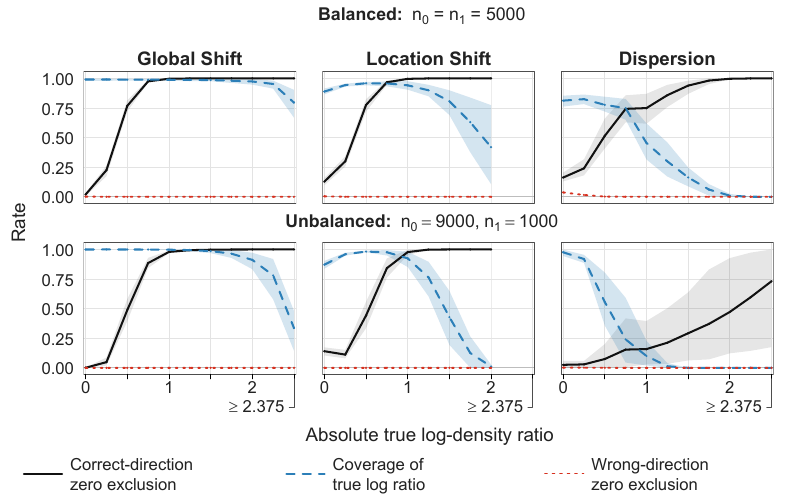}
    \caption{\rnew{
    Pointwise evaluation of the credible intervals in the 20-dimensional simulations as a function of the absolute true log-density ratio.
    \rnew{Columns show Global Shift, Location Shift, and Dispersion; rows show balanced sample sizes} $(n_0,n_1)=(5000,5000)$ \rnew{and unbalanced sample sizes} $(n_0,n_1)=(9000,1000)$.
    The black line shows the rate at which zero is excluded in the correct direction, the blue dashed line shows the coverage rate of the true log-density ratio, and the red dotted line shows the rate at which zero is excluded in the wrong direction.
    Correct-direction exclusion means that the lower endpoint is above zero when the true log-density ratio is positive, or that the upper endpoint is below zero when the true log-density ratio is negative.
    \rnew{Lines average 50 simulation repeats; shading shows the corresponding empirical 10th and 90th percentiles.}
    The two groups receive equal total weight when the pointwise rates are calculated.
    }}
    \label{fig:20d-pointwise-evaluation}
\end{figure}

\rnew{
Given the undercoverage observed in some 20-dimensional settings,  we need to clarify what information the credible intervals can provide beyond coverage of the exact log-density-ratio magnitude.
To this end, Figure~\ref{fig:20d-pointwise-evaluation} compares exact-value coverage with zero exclusion as a function of the absolute true log-density ratio.
The probability of excluding zero in the correct direction generally increases as the true log-density ratio moves farther from zero, while exclusion in the wrong direction remains rare.
In particular, under the Dispersion scenario, an interval may correctly exclude zero even when it does not cover the exact magnitude of the true log-density ratio.
Thus, the intervals can retain useful information about the existence and direction of a pointwise two-sample difference even when exact-value coverage is imperfect.
}

\begin{table}[htb]
\centering
\caption{A comparison of the MSEs and their standard errors in the 20-dimensional scenarios.
}\label{table: MSE(multi)}
\begin{tabular}{lcccccc}\toprule
    &\multicolumn{2}{c}{\textbf{Global Shift}}
    &\multicolumn{2}{c}{\textbf{Location Shift}}
    &\multicolumn{2}{c}{\textbf{Dispersion}}
    \\
    \cmidrule(r){2-3}\cmidrule(r){4-5}\cmidrule(r){6-7}
    $n_0 / (n_0 + n_1)$ &0.5&0.9&0.5&0.9&0.5&0.9\\\midrule
GB &\rnew{0.049} &\rnew{0.090} &\rnew{0.057} &\rnew{0.097} &\rnew{0.128} &\rnew{0.210} \\
&\rnew{(0.001)} &\rnew{(0.002)} &\rnew{(0.001)} &\rnew{(0.002)} &\rnew{(0.003)} &\rnew{(0.005)} \\
FS &\rnew{0.049} &\rnew{0.097} &\rnew{0.059} &\rnew{0.102} &\rnew{0.137} &\rnew{0.213} \\
&\rnew{(0.001)} &\rnew{(0.002)} &\rnew{(0.001)} &\rnew{(0.002)} &\rnew{(0.003)} &\rnew{(0.005)} \\
BAT &\rnew{0.031} &\rnew{0.059} &\rnew{0.055} &\rnew{0.098} &\rnew{0.165} &\rnew{0.339} \\
&\rnew{(0.001)} &\rnew{(0.002)} &\rnew{(0.001)} &\rnew{(0.002)} &\rnew{(0.004)} &\rnew{(0.008)} \\
DRT (AdaBoost) &\rnew{0.061} &\rnew{0.139} &\rnew{0.070} &\rnew{0.173} &\rnew{0.137} &\rnew{0.264} \\
&\rnew{(0.002)} &\rnew{(0.003)} &\rnew{(0.001)} &\rnew{(0.003)} &\rnew{(0.003)} &\rnew{(0.005)} \\
CDC (AdaBoost) &\rnew{0.488} &\rnew{1.208} &\rnew{0.160} &\rnew{0.600} &\rnew{0.534} &\rnew{1.706} \\
&\rnew{(0.022)} &\rnew{(0.044)} &\rnew{(0.003)} &\rnew{(0.020)} &\rnew{(0.022)} &\rnew{(0.055)} \\
KLIEP &\rnew{0.566} &\rnew{0.593} &\rnew{0.616} &\rnew{0.614} &\rnew{0.547} &\rnew{0.548} \\
&\rnew{(0.012)} &\rnew{(0.014)} &\rnew{(0.005)} &\rnew{(0.008)} &\rnew{(0.001)} &\rnew{(0.001)} \\
uLSIF &\rnew{0.602} &\rnew{0.617} &\rnew{3.422} &\rnew{3.678} &\rnew{0.546} &\rnew{0.547} \\
&\rnew{(0.009)} &\rnew{(0.006)} &\rnew{(0.081)} &\rnew{(0.068)} &\rnew{(0.001)} &\rnew{(0.001)} \\
            \bottomrule
\end{tabular}%
\begin{tablenotes}
\small
\item Notes: The methods are GB (gradient boosting), FS (forward stagewise), BAT (Bayesian additive trees), DRT \nanew{and CDC} (density-ratio trick and \nanew{calibrated discriminative classifier} based on AdaBoost), and the two kernel-based methods KLIEP and uLSIF.
\end{tablenotes}
\end{table} 

\vspace{-1.5em}
\Needspace{4\baselineskip}\section{Assessing Synthesized Microbiome Compositions}
\label{sec:real_data}
\vspace{-0.5em}

\rnew{Recent generative models imitate the sampling mechanisms of complex biomedical data.} In particular, several approaches have been proposed for generating microbiome compositions \citep{sankaran2025semisynthetic}. \rnew{A common need is to assess their synthetic samples against observed data.}
We apply our two-sample comparison framework to synthetic samples simulated from different generative models and the corresponding observed samples, thereby assessing the quality of data generation in these models. 

The real microbiome data are obtained from the open-source Curated Metagenomic Data \citep{pasolli2017accessible}, available as the R package \texttt{curatedMetagenomicData}. \rnew{We focus on species-level relative abundance data. First, we retain observations whose total relative abundance is at least $99\%$. We then retain taxa whose mean relative abundance across these observations is at least $0.1\%$. Among observations satisfying the first criterion, we retain those whose mean abundance over the selected taxa is at least the corresponding average across all observations in the original collection. This procedure retains 123 taxa. Finally, we renormalize each retained observation over these taxa so that its relative abundances sum to one.}

\rnew{We apply the above preprocessing rule to \texttt{curatedMetagenomicData} and obtain pooled microbiome observations from IBDMDB \citep{ibdmdb2019}, the longitudinal prediabetes study \citep{zhou2019longitudinal}, and the studies of \citet{hall2017novel} and \citet{hannigan2018diagnostic}.}
The samples are in fact collected \rnew{separately in these different studies, and at} multiple time points for each subject. \rnew{Both the subgroup structure and} the longitudinal nature of the data are important for scientific analysis.
Here we ignore these aspects for demonstration and focus only on assessing how well generative models can approximate the marginal distribution of the samples. We split the data into training ($n=1641$) and testing ($n=411$) sets. 

\rnew{We fit the generative models to the training data and compare their synthetic samples with the real test data.} We evaluate various popular generative models, including both classical parametric models and state-of-the-art deep neural network-based models. The parametric models considered are the Dirichlet model \citep{holmes2012,larosa2012} and the Dirichlet tree model \citep{dennis1991,wang2017dirichlet,mao2022dirichlet}. The nonparametric generators based on neural networks include continuous normalizing flows via conditional flow matching \citep[ICFM]{lipman2022flow,tong2023improving}%
, and a generative adversarial network model called MB-GAN \citep{rong2021mb}%
, which is tailored for microbiome data generation. Our objective here is not to argue which model is better than others, as such judgments are context-dependent and require more careful tuning of each model. Rather, we aim to provide an example that density ratio-based two-sample comparison can provide insights into how generative models can be gauged given the observations that they generate.

\begin{figure}[ht!]
    \centering
    \includegraphics[width=\linewidth]{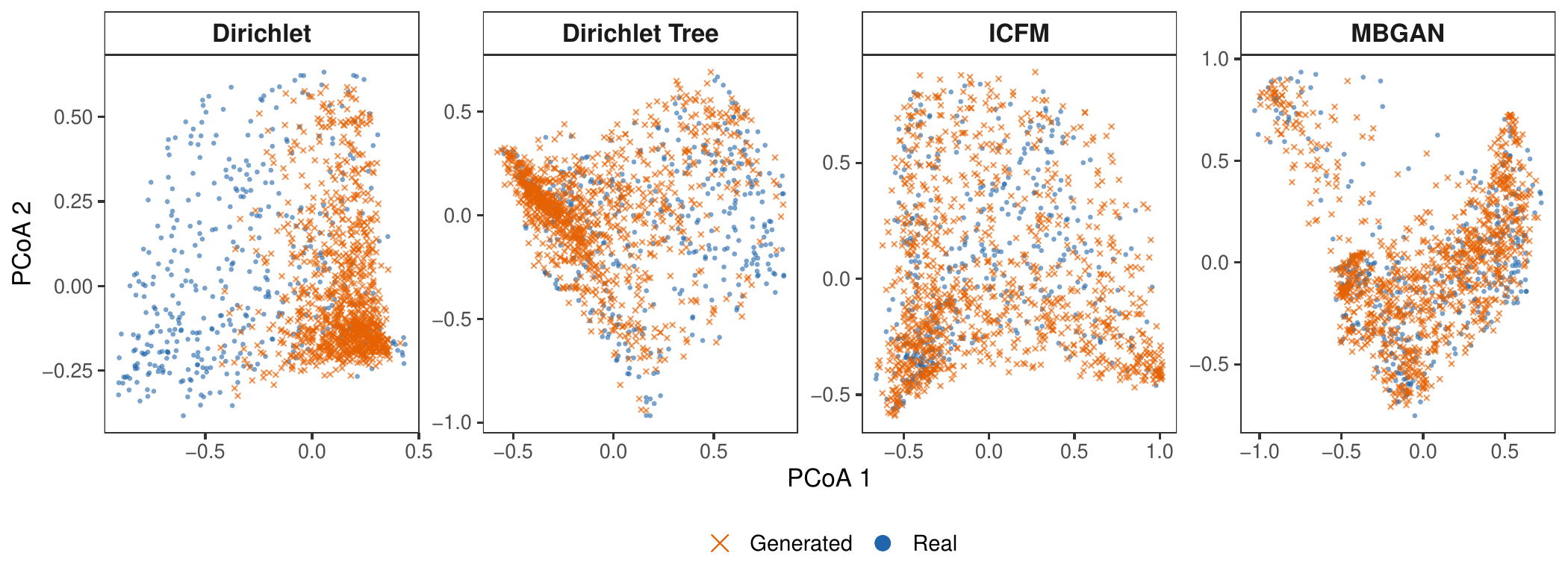}
    \caption{Principal Coordinates Analysis (PCoA) plots based on Bray-Curtis distances among \yxnew{test} observations.
    \rnew{Blue and orange points represent real and generated samples, respectively.} 
   }
    \label{fig:PCoA_raw}
\end{figure}

The PCoA plots based on the Bray-Curtis distance \citep{bray1957ordination} in Figure \ref{fig:PCoA_raw} \rnew{provide a low-dimensional comparison of the generated and test samples} \yxnew{(see Figure~\ref{fig:PCoA_raw_train} for the training sample)}. \rnew{These plots show that  parametric generators such as the Dirichlet model may have less coverage of the support than the real sample, while the two neural network-based approaches perform better in this regard.} \rnew{However, these visualizations offer limited differentiation between the Dirichlet tree and the two neural network-based generators.}

\begin{figure}[t]
  \centering
\includegraphics[width=0.7\textwidth]{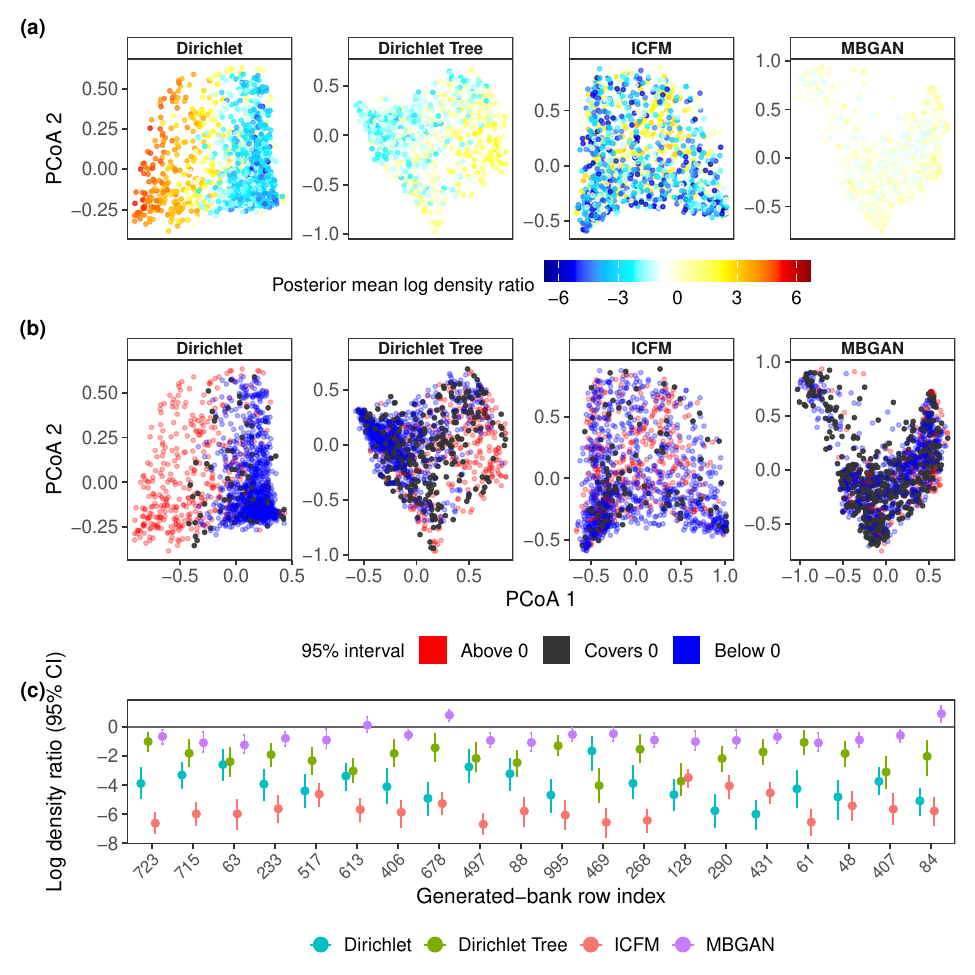}
  \caption{For the test data: (a) PCoA plots (x-axis and y-axis are the first and second principal components) colored by the density ratio estimates of the testing sample density over the generated sample density, and (b) whether the 2.5\% and 97.5\% credible bands cover 0; (c) For \rnew{the top 20 observations} with the largest absolute log-density ratio averaged over all methods, the point estimates and the credible intervals for each generative model. \yxnew{See the training data equivalent plot in Figure~\ref{fig:RDA_DRE_train}.}}
  \label{fig:RDA_DRE_test}
\end{figure}

\rnew{We apply generalized Bayesian additive trees to obtain posterior density-ratio samples for each pair of real and synthetic samples.} \rnew{We compute the posterior mean and pointwise 95\% credible band for the density ratio}, as shown in \yxnew{Figure~\ref{fig:RDA_DRE_test}(a) for the testing sample \rnew{versus} the synthetic sample (Figure~\ref{fig:RDA_DRE_train}(a) for the training sample \rnew{versus} the synthetic sample)}. The color indicates the estimated (log) density ratio. Figure~\ref{fig:RDA_DRE_test}(b) and Figure~\ref{fig:RDA_DRE_train}(b) \rnew{visualize observation-level credible interval coverage through PCoA, showing whether each interval is below, covers, or is above 0.} Figure~\ref{fig:RDA_DRE_test}(c) and Figure~\ref{fig:RDA_DRE_train}(c) provide the credible intervals for the top 20 observations with the largest absolute log-density ratio values. \rnew{Supplementary Materials~\ref{appendix: additional figures} provides detailed plots of the credible bands for all samples.} Recall that a log-density ratio of zero indicates that the generated and real samples are equally represented at the distributional level, which is the defining property of an ideal generative model. The estimated density ratios in both Figures \ref{fig:RDA_DRE_train} and \ref{fig:RDA_DRE_test} provide a clear contrast of the generation quality compared to the true sample. Compared with other generative models, the estimated log-density ratio for the MB-GAN-generated sample versus the real sample is closer to 0 across the support, especially in the test data.
\rnew{As shown in Figure~\ref{fig:RDA_DRE_test}(b), the credible bands provided by our proposed method quantify uncertainty in the density ratio. For MB-GAN, the 95\% pointwise credible intervals for the log-density ratio include zero for 38.1\% of synthetic observations and 50.4\% of real test observations. MB-GAN contains the largest fraction of both real and synthetic observations whose pointwise CIs contain 0 compared to other generative models.}

\yxnew{\rnew{To check for overfitting in the high-dimensional microbiome setting, we conduct a null experiment that randomly divides the training observations into two equal groups treated as samples from} \(P\) and \(Q\). Because both groups were drawn from the same empirical population, the true log-density ratio is zero. We fitted BAT to these training groups and evaluated the posterior log-density ratio on the independent test set. All 411 test samples' pointwise 95\% posterior credible intervals contained zero (see Figure~\ref{fig:null_test}). \rnew{This provides evidence that BAT does not identify spurious distributional differences on held-out high-dimensional observations in this null setting, compared with} Figure~\ref{fig:RDA_DRE_test}.}

\vspace{-1em}
\section{Conclusion}\label{sec:conclusion}
\vspace{-0.5em}

\rnew{We have introduced density ratio learning between two i.i.d.\ samples using additive tree models and a new loss function, the balancing loss.}
\rnew{It provides effective boosting algorithms for functional (point) estimates and a convenient means for}
attaining uncertainty quantification, which can be critical in challenging problems with limited sample size or high-dimensionality. 

\rnew{Additive tree models also have limitations.}
\rnew{Their axis-aligned partitions can sometimes capture high-order interactions less effectively}, and they generally do not perform well with extremely high-dimensional data, i.e., those involving more than a few hundred dimensions.
\rnew{High-dimensional structures can also yield poor overlap between two samples, which can result in unstable performance of the proposed algorithms.}
\rnew{In such scenarios, an effective strategy couples tree-based approaches with learned data representations.}
Examples of such representations might include low-dimensional subspace assumptions (e.g., manifolds) or other non-linear embeddings through state-of-the-art neural network-based encoders. 
\rnew{Another strategy is to address the poor overlap by comparing one group with a mixture of the two groups as proposed in \citet{xu2025distributional}, though this requires reformulation of our inferential problem.}
\rnew{We argue, however, that even in such cases, learning the density ratio between two samples is a worthwhile objective and will typically be easier than learning the density of each sample {\em per se}.}

\vspace{-2em}

\section*{Software}
\vspace{-1em}

\noindent
\hidetext{
\rnew{We have implemented an R package {\tt BATTS} for the proposed boosting and Bayesian additive tree algorithms, and it is available at \href{https://github.com/nawaya040/BATTS}{https://github.com/nawaya040/BATTS}. The code to reproduce the simulation studies in Section~\ref{sec: numerical experiments} is available at \\\href{https://github.com/nawaya040/BATTS_simulation_study}{https://github.com/nawaya040/BATTS\_simulation\_study}.}
The case study reproducible code from data preprocessing, generative model training, and {\tt BATTS} evaluation and visualization is available at \href{https://github.com/yuliangxu/TwoSample}{https://github.com/yuliangxu/TwoSample}.}

\vspace{-2em}

\section*{Acknowledgment}
\vspace{-0.5em}

\noindent
\hidetext{
This research is funded in part by NIGMS grant R01-GM135440 and NSF grants DMS-2152999 and DMS-2531539. Part of the research was completed when LM and YX were at Duke University.
NA's research is supported by the Japan Society for the Promotion of Science KAKENHI (grant numbers: 25K16618, 24H00142, and 25H00546).
Portions of the manuscript were prepared with the assistance of ChatGPT (from GPT-4.1 to 6.0) and Claude Code (Opus 5.5) for English-language editing, manuscript revision, and coding.
} 
\vspace{-2em}

\section*{Data Availability Statement}
\vspace{-1em}

\noindent
The real data, curated metagenomic data \citep{pasolli2017accessible}, used in Section~\ref{sec:real_data} are openly available as an R package \texttt{curatedMetagenomicData} on GitHub and Bioconductor. The instruction and documentation are at \url{https://waldronlab.io/curatedMetagenomicData/}, DOI: 10.18129/B9.bioc.curatedMetagenomicData.

\vspace{-1.5em}

\begingroup
\setstretch{1.5}
\setlength{\bibsep}{2pt plus 1pt minus 1pt}
\bibliographystyle{agsm} 
\bibliography{references} 
\endgroup

\end{document}

\clearpage
\appendix
\setcounter{figure}{0}
\setcounter{table}{0}
\renewcommand{\thefigure}{S\arabic{figure}}
\renewcommand{\thetable}{S\arabic{table}}
\renewcommand{\bibliographystyle}[1]{}
\renewcommand{\bibliography}[1]{}

\begin{center}
  {\Large\bfseries Supplementary Materials for}\\[0.4em]
  {\Large\bfseries ``Two-sample Comparison through Additive Tree Models\\[0.2em]
   for Density Ratios''}
\end{center}
\vspace{1em}

\documentclass[12pt]{article}
\usepackage{amsmath, amssymb, amscd, amsthm, amsfonts}
\usepackage{graphicx}

\usepackage{here}
\usepackage[affil-it]{authblk}
\usepackage{setspace}
\usepackage{subcaption}

\usepackage[utf8]{inputenc}
\usepackage{booktabs}

\usepackage[round]{natbib}
\usepackage{bm}

\usepackage{xr-hyper}
\usepackage{hyperref}
\externaldocument[][nocite]{JASA_submission_R2/main_JASA_Aug10_2026}
  [main_JASA_Aug10_2026.pdf]

\usepackage{threeparttable} 

\oddsidemargin 0pt
\evensidemargin 0pt
\marginparwidth 40pt
\marginparsep 10pt
\topmargin -20pt
\headsep 10pt
\textheight 8.7in
\textwidth 6.65in
\linespread{1.2}

\newif\ifhidetext
\hidetexttrue 
\newcommand{\hidetext}[1]{%
  \ifhidetext
    [hidden for blinding]
  \else
    #1
  \fi
}

\newcommand{\hideauthor}[1]{%
  \ifhidetext
    \author{}
  \else
    #1
  \fi
}

\renewcommand{\thesection}{\Alph{section}}

\title{Supplementary Materials for ``{Two-sample Comparison through Additive Tree Models for Density Ratios}''}
\hideauthor{
\author[1]{Naoki Awaya}
\author[2]{Yuliang Xu}
\author[2]{Li Ma}
\affil[1]{School of Political Science and Economics, Waseda University}
\affil[2]{Department of Statistics \& Data Science Institute, University of Chicago}
}

\date{}

\usepackage{xcolor}
\newcommand\yxnew[1]{{\color{blue} {#1}}} 
\newcommand\yxcm[1]{{\color{blue} {(#1)}}} 

\newcommand\nanew[1]{#1} 
\newcommand\rnew[1]{{\color{blue} {#1}}} 

\newcommand{\argmax}{\mathop{\rm arg~max}\limits}
\newcommand{\argmin}{\mathop{\rm arg~min}\limits}
\newcommand\real{\mathbb{R}}

\newtheorem{thm}{Theorem}[section]
\newtheorem{lem}{Lemma}[section]
\newtheorem{cor}{Corollary}[section]
\newtheorem{prop}{Proposition}[section]

\usepackage{amsmath}               
  {
      \theoremstyle{plain}
      \newtheorem{assumption}{Assumption}
  }

\newcommand{\1}{\mathbf{1}}

\newcommand{\x}{\mathbf{x}}
\newcommand{\y}{\mathbf{y}}
\newcommand{\h}{\mathbf{h}}

\newcommand{\Q}{\mathbf{Q}}

\newcommand{\C}{\mathcal{C}}
\newcommand{\A}{\mathcal{A}}
\newcommand{\B}{\mathcal{B}}
\newcommand{\T}{\mathcal{T}}
\newcommand{\Z}{\mathbb{Z}}
\newcommand{\Polya}{P\'{o}lya\ }
\newcommand{\F}{\mathbf{F}}
\newcommand{\tF}{\tilde{\mathbf{F}}}
\newcommand{\G}{\mathbf{G}}
\newcommand{\tG}{\tilde{\mathbf{G}}}
\newcommand{\cP}{\mathcal{P}}

\newcommand{\D}{\mathbf{D}}

\newcommand{\R}{\mathbb{R}}
\newcommand{\E}{\mathbb{E}}

\newcommand{\Var}{\mathrm{Var}}
\newcommand{\Cov}{\mathrm{Cov}}

\newcommand{\Bspace}{\mathcal{B}^2}

\setcounter{figure}{0}
\renewcommand{\thefigure}{S\arabic{figure}}

\begin{document}
\maketitle

\doublespacing

\section{Proofs}
\label{appendix: proofs}
\subsection{Proof of Proposition~1}

We first fix the finite tree partition $T_k$.
After adding the piecewise constant function $f_k$, the loss is given by
\[
    l(w)
    =
    \sum_{A \in T_k}
    \left[
        \exp(-\beta_k(A))
        \int_A w^{-1}_{k-1} p\, d \mu
        +
        \exp(\beta_k(A))
        \int_A w_{k-1} q\, d \mu  
    \right]
\]
By the relationship between the arithmetic mean and the geometric mean, each summand is bounded below by
\[
    2\sqrt{
    \int_A w^{-1}_{k-1} p\, d \mu
    \int_A w_{k-1} q\, d \mu  
    },
\]
and the equality is attained if
\[
    \exp(-\beta_k(A))
    \int_A w^{-1}_{k-1} p\, d \mu
    =
    \exp(\beta_k(A))
    \int_A w_{k-1} q\, d \mu ,
\]
and this condition yields the expression of the optimal $\beta_k(A)$.
Furthermore, when the lower bound is attained for all $A \in T_k$, the loss can be expressed as
\[
    l(w)
    =
    2
    \sum_{A \in T_k}
    \sqrt{
    \int_A w^{-1}_{k-1} p\, d \mu
    \int_A w_{k-1} q\, d \mu  
    },
\]
and minimizing it with respect to $T_k$ is equivalent to maximizing the discrete Hellinger distance given in the claim.
\qed

Also, we can derive the following corollary by considering the case in which the $T_k$ is the root tree.

\begin{cor}
    Let $w$ be the current estimate of the balancing function and $c$ be the constant defined as $c = \sqrt{\mathbb{E}_p[w^{-1}]/\mathbb{E}_q[w]}$ and thus satisfies
    \[
        \mathbb{E}_p[(c w)^{-1}] = \mathbb{E}_q[(c w)].
    \]
    Then the balancing loss is improved as
    \[
        l(w) \geq l(cw).
    \]
\end{cor}

\subsection{\rnew{Derivation of the Regularized Prior for the Node Parameters}}
\label{appendix: prior derivation}
According to \cite{whitmore1978normalizing}, when $W \sim \mathrm{InverseGauss} (1, \theta)$ and $\theta$ is large, the distribution of the following variable is approximated by the standard normal distribution:
\[
    \frac{1}{2 \sqrt{\theta}}
    +
    \sqrt{\theta} \log W.
\]
Hence, when $\gamma_k(A) = \exp(\beta_k(A))$ follows $\mathrm{InverseGauss} (1, K \lambda_0)$, the leaf node parameter $\beta_k(A)$ approximately follows
\[
    N  
    \left(
        \rnew{- \frac{1}{2 K \lambda_0}},
        \frac{1}{K \lambda_0}
    \right).
\]
This observation implies that, under the alternating prior specification
\[
    \gamma_k(A)
    \sim
    \mathrm{InverseGauss} (1, K \lambda_0)\ (k: \text{odd}),\
    \gamma_k(A)^{-1}
    \sim
    \mathrm{InverseGauss} (1, K \lambda_0)\ (k: \text{even}),
\]
the distribution of $\log w(x)$ for fixed $x$ is approximated by $N(0, \lambda_0^{-1})$.

\section{\rnew{Details of the Posterior Sampling for the Bayesian Additive Tree Model}}
\label{appendix: Details of the MCMC algorithm}

\subsection{\rnew{MCMC Algorithm}}
The Bayesian additive model includes three types of unknown variables: (i) tree partition structures $T_1,\dots,T_K$, (ii) leaf node parameters $\beta_1,\dots,\beta_K$, where $\beta_k$ denotes the collection of the parameters for the $k$th basis function, and (iii) the temperature $\tau$. 
Their posterior distributions are estimated with the following Metropolis-within-Gibbs algorithm:

\begin{enumerate}
    \item For $k=1,\dots,K$, 
    \begin{enumerate}
        \item Compute the residual $\log w_{-k} = \sum_{l \neq k}f_l$.
        \item Update $T_k \mid w_{-k}, \tau$.
        \item Update $\beta_k \mid w_{-k},T_k, \tau$.
    \end{enumerate}
    \item Update $\tau \mid w$, where $w$ is the current balancing weight.
\end{enumerate}
Updating the node parameters $\beta_k$ and the temperature $\tau$ is straightforward due to their conjugacy.
The algorithm to update $T_k$ with the Metropolis-Hastings algorithm is detailed in the next section. 

\subsection{\rnew{Updating Trees}}
In this section, we update the $k$th tree, denoted by $T_k$, conditional on the other parameter values, using proposals originally introduced for Bayesian CART \citep{chipman1998bayesian}. The details follow \cite{chipman2010bart, kapelner2016bartmachine}.

In each iteration of the MCMC, we propose a new tree structure $T^*_k$ by modifying the current tree $T_k$ via one of the three moves, GROW, PRUNE, and CHANGE, defined as follows:
\begin{description}
    \item[GROW: ] Randomly select one leaf node and split it to generate two child nodes.
    \item[PRUNE: ] Randomly select one second-generation internal node (2GI), that is, a node with two child nodes that are terminal, and remove these child nodes.
    \item[CHANGE: ] Randomly select one 2GI node and change the splitting rules by sampling a new rule from the prior distribution.
\end{description}
One of the moves is selected stochastically according to the fixed probabilities $p(\mathrm{GROW})$, $p(\mathrm{PRUNE})$, and $p(\mathrm{CHANGE})$, and they are all set to 1/3 in our numerical experiments. 

Let $T^*_k$ denote the new proposed tree structure.
Then, the proposal is accepted with probability $\min \{\alpha, 1\}$, where
\[
    \alpha
    =
    \underbrace{
    \frac{p(T^*_k)}{p(T_k)}
    }_{\text{Prior}}
    \underbrace{
    \frac{p(T^*_k \to T_k)}{p(T_k \to T^*_k)}
    }_{\text{Transition}}
        \underbrace{
    \frac{p(\mathrm{data} \mid T^*_k,\{w_{-k}(x_i)\}^n_{i=1})}{p(\mathrm{data} \mid T_k, \{w_{-k}(x_i)\}^n_{i=1})}
    }_{\text{Loss-based likelihood}}
\]
and $\{w_{-k}(x_i)\}^n_{i=1}$ are the residuals.
If a node $A$ is selected for updating $T_k$, then the ratios corresponding to the priors and the transition probabilities for the three moves are given as
\begin{align*}
    \frac{p(T^*_k)}{p(T_k)}
    \frac{p(T^*_k \to T_k)}{p(T_k \to T^*_k)}
    =
    \begin{cases}
    \frac{
        a_T
        \left(
            1 - a_T (2 + \mathrm{depth}(A))^{-b_T}
        \right)^2
    }{
        \left(
            1 + \mathrm{depth}(A)           
        \right)^{b_T}
        -
        a_T
    }
    \frac{\text{\# leaf nodes in }T_k}{\text{\# 2GI nodes in }T^*_k}\ &(\text{GROW}),\\
    \frac{
               \left(
            1 + \mathrm{depth}(A)           
        \right)^{b_T}
        -
        a_T
    }{
        a_T
        \left(
            1 - a_T (2 + \mathrm{depth}(A))^{- b_T}
        \right)^2
    }
    \frac{\text{\# 2GI nodes in }T_k}{\text{\# leaf nodes in }T^*_k}\ &(\text{PRUNE})\\
    1\ &(\text{CHANGE}).
    \end{cases}
\end{align*}
(see \cite{kapelner2016bartmachine} for further details.)

For calculating the likelihood ratios, we need to note that we can analytically evaluate the likelihood with the leaf node parameters integrated out due to conjugacy as follows:
\[
    \int p(\beta_k) L_{n,\tau} (w)\, d \beta_k
    \propto 
    \prod_{A \in T_k}
    L_{k,A},
\]
where $p(\beta_k)$ is the joint prior density of the leaf node parameters, and 
\[
    L_{k,A}
    =
    \sqrt{
        \frac{\lambda_k(A)}{\lambda'_k(A)}
    }
    \exp
    \left(
        \frac{\lambda_k(A)}{\mu_k(A)}
        -
        \frac{\lambda'_k(A)}{\mu'_k(A)}
    \right).
\]
Hence, for the GROW move in which the node $A$ is split into $A_l$ and $A_r$, the likelihood ratio is given as
\[
    \frac{L_{k, A_l}L_{k, A_r}}{L_{k,A}},
\]
while in the PRUNE move, which removes $A_l$ and $A_r$, the ratio is the inverse.
The CHANGE move generates two new children $A^*_l$ and $A^*_r$, so the likelihood ratio is given as
\[
    \frac{L_{k,A^*_l}L_{k,A^*_r}}{L_{k,A_l}L_{k,A_r}}.
\]

\section{\rnew{Details of Experiments}}
\label{appendix: details of experiments}

\subsection{\rnew{2-dimensional Numerical Experiments}}
\begin{description}
    \item[1. Global Shift]
    $
        X_0 
        \sim
        N
        \left(
            \mu^{(0)},
            I_2
        \right),\ 
        X_1 
        \sim
        N
        \left(
           \mu^{(1)},
            I_2
        \right)$,
    where $\mu^{(0)} = (-0.5, -0.5)'$ and \rnew{$\mu^{(1)} = (0, 0)'$.}
    \vspace{-1em}
    \item[2. Local Shift]
    \[
    X_0 
    \sim
    \frac{1}{5}
        N(\mu_1, \Sigma_1)
    +
    \frac{1}{5}
    \sum^5_{k=2}
        N(\mu_k, \Sigma_k),\ 
    X_1 
    \sim
    \frac{1}{5}
        \rnew{N(\mu_5 + \delta, \Sigma_5)}
    +
    \frac{1}{5}
    \rnew{\sum^4_{k=1}}
        N(\mu_k, \Sigma_k),
    \]
    \vspace{-3em}

\noindent
    where $\delta  = (0, 1)'$.
    \vspace{-1em}
    \item[3. Local Dispersion Difference]
    \[
    X_0 \sim \frac{1}{3}\sum_{k=1}^{2}N(\mu_k,\Sigma_k)
        +\frac{1}{3}N(\mu_3,D\Sigma_3D^{\mathsf T}),\qquad
    X_1 \sim \frac{1}{3}\sum_{k=1}^{3}N(\mu_k,\Sigma_k),
    \]
    \vspace{-3em}

    \noindent
    where $D = \mathrm{diag}(0.6,1)$.
\end{description}
The parameters for simulating the data sets are specified as follows:

\begin{description}
    \item[2. Local Shift]
$\mu_1 = (9.0, 9.9)'$, $\mu_2 = (-2.5, 1.4)'$, $\mu_3 = (-2.3, -9.7)'$, $\mu_4 = (3.4, 5.9)'$, $\mu_5 = (5.8, -9.5)'$, $(\Sigma_1(1,1), \Sigma_1(1,2), \Sigma_1(2,2)) = (2.9, 0.5, 1.1)$, $(\Sigma_2(1,1), \Sigma_2(1,2), \Sigma_2(2,2)) = (1.2, -0.6, 2.8)$, \\
$(\Sigma_3(1,1), \Sigma_3(1,2), \Sigma_3(2,2)) = (2.3, -1.0, 1.7)$, $(\Sigma_4(1,1), \Sigma_4(1,2), \Sigma_4(2,2)) = (1.1, -0.4, 2.9)$, \\
and $(\Sigma_5(1,1), \Sigma_5(1,2), \Sigma_5(2,2)) = (3.0, 0.2, 1.0)$.
    \item[3. Local Dispersion Difference]
    $\mu_1 = (1.9, -7.2)'$, $\mu_2 = (-2.3, -1.5)'$, $\mu_3 = (7.5, -3.1)'$, $(\Sigma_1(1,1), \Sigma_1(1,2), \Sigma_1(2,2)) = (1.0, -0.4, 0.8)$,
$(\Sigma_2(1,1), \Sigma_2(1,2), \Sigma_2(2,2)) = (1.0, 0, 3.0)$, and $(\Sigma_3(1,1), \Sigma_3(1,2), \Sigma_3(2,2)) = (2.9, 0, 1.1)$.
\end{description}

\subsection{\rnew{20-dimensional Experiments}}
$\mu_1$ is set to $(-0.5, 0, 0, 0)'$ in the Location Shift scenario and $(0,0,0,0)'$ in the Dispersion Difference scenario.
The other parameters are fixed as $\mu_2 = (0,0,0,0)'$, $\Sigma_1 = I_4$, and $\Sigma_2 = 4 I_4$ in both scenarios. 

\subsection{\rnew{Notes on the Implementation of the Density Ratio Estimators}}
For boosting, we assess both the FS and the GB algorithms. The learning rate $\nu$ is set to 0.01, and the maximum resolution of each tree is set to 4. The optimal number of trees is selected by five-fold cross-validation with the maximum number allowed set to 1000.  In both types of the boosting algorithms, we introduce 31 equally spaced cut points in each dimension.
\rnew{As in AdaBoost, each tree is constructed from a random half of the training observations (a subsampling fraction of 0.5), drawn as in the {\tt bag.frac} option of {\tt ada}: one observation is selected from each group, and the remaining observations are selected uniformly without replacement regardless of group. The weights of all training observations are updated after each tree.}
\rnew{For AdaBoost, we use the \texttt{ada} function from the \texttt{ada} package \citep{ada_package} with Real AdaBoost (\texttt{type = "real"}), a learning rate of 0.01, a subsampling fraction of 0.5 (the package default), and a maximum of 1000 trees. Each tree is grown to depth 4 without cost-complexity pruning or a minimum node size (\texttt{rpart.control(maxdepth = 4, cp = -1, minsplit = 0)}), matching the maximum resolution of the trees in our boosting algorithms. The number of trees is selected by five-fold cross-validation with folds stratified by group, minimizing the exponential loss averaged over the folds.}
For the generalized Bayesian additive tree model, we set the size of the ensemble to 200, following the standard implementation of BART \citep{chipman2010bart}.
The MCMC sampler is initialized with trees containing only root nodes, and the size of the burn-in and the posterior sample is set to 2000 and 1000, respectively. 

\rnew{Additionally, in the boosting algorithms, we exclude candidate splits that would produce a child node containing no observations from one of the two groups among the observations used to construct the tree, which avoids division by zero in computing the optimal values of the leaf node parameters.}

\rnew{For the calibration method \citep{cranmer2015approximating}, we apply the kernel density estimator implemented by the {\tt density} function in {\tt R} separately to the AdaBoost scores in the two groups, after subtracting the empirical log class-odds offset $\log(n_1/n_0)$. The calibrated log-density ratio is $\log\{\widehat g_0(s)/\widehat g_1(s)\}$, where $\widehat g_0$ and $\widehat g_1$ are the estimated score densities for groups 0 and 1. Supplementary Materials~\ref{sec:cdc-calibration} gives the numerical evaluation details.}

For the KLIEP and uLSIF, we use the functions \texttt{KLIEP} and \texttt{uLSIF} from the R package \texttt{densratio} \citep{densratio_package} with the default parameter values.
It should be noted that we detected an error in the source code for the \texttt{KLIEP} function
\footnote{
    The estimated density ratios are not correctly computed on observations in a test data set in the original package.
}
, so we used a fixed version that is available at \hidetext{\url{https://github.com/nawaya040/densratio}}.

We found that the compared methods can output infinite values. Such values are omitted in computing the means and the standard errors of the results. 
We note that such phenomena were not observed for the proposed methods.

\newpage 

\section{\rnew{Additional Numerical Experiment Results}}
\label{appendix: additional figures}

\subsection{\rnew{Relationship between the Exponential and Balancing Losses}}
\label{appendix: loss relationship}
\rnew{
We describe how the exponential loss used by AdaBoost relates to the balancing loss when the two sample sizes differ.
Here we assign $y=1$ to the $n_0$ observations $x^{(0)}_1,\dots,x^{(0)}_{n_0}$ from $p$ and $y=-1$ to the $n_1$ observations $x^{(1)}_1,\dots,x^{(1)}_{n_1}$ from $q$.
The empirical exponential loss of a function $F$ is
\[
    \frac{1}{n_0 + n_1}
    \left(
        \sum_{i=1}^{n_0} e^{-F(x^{(0)}_i)}
        +
        \sum_{j=1}^{n_1} e^{F(x^{(1)}_j)}
    \right).
\]
Let $c=\frac{1}{2}\log(n_0/n_1)$.
If the tree ensemble is initialized at the constant $c$, or equivalently, if each observation is weighted inversely proportional to the size of its group, the loss function becomes
\begin{align*}
    &\frac{1}{n_0 + n_1}
    \left(
        \sum_{i=1}^{n_0} e^{-\{F(x^{(0)}_i)+c\}}
        +
        \sum_{j=1}^{n_1} e^{F(x^{(1)}_j)+c}
    \right)\\
    &\propto
    \frac{1}{n_0}\sum_{i=1}^{n_0} e^{-F(x^{(0)}_i)}
    +
    \frac{1}{n_1}\sum_{j=1}^{n_1} e^{F(x^{(1)}_j)},
\end{align*}
which is the balancing loss of $w=e^{F}$.
Hence, minimizing the exponential loss of $F+c$ is equivalent to minimizing the balancing loss of $w=e^{F}$; in this case, $F$ estimates half the log density ratio, whereas $F+c$ estimates half the log posterior odds.
}

\subsection{\rnew{Comparison with the Density Ratio Trick in the One-dimensional Scenarios}}
\label{appendix: dr trick 1d}
\nanew{
We perform a simple experiment to discuss how the bias can be introduced by the density ratio trick based on classification methods, if the prior information on the sample size is not effectively incorporated into the learning algorithms.

\rnew{As in Section~\ref{sec: uncertainty calibration}, we simulate data sets from the two Gaussian distributions, $N(0, 1)$ and $N(1, 2.25)$, and compare two algorithms: (i) the density-ratio trick based on Real AdaBoost, implemented using the {\tt ada} function from the {\tt ada} package \citep{ada_package}, and (ii) the gradient boosting algorithm that optimizes the proposed balancing loss.}
\rnew{For both algorithms, the maximum tree depth and ensemble size are set to 2 and 100, respectively, the learning rate is 0.01, and each tree is fitted to a random half of the observations (a subsampling fraction of 0.5).}
\rnew{This small, fixed ensemble is used to isolate the effect of the initial value; the shrinkage toward the initial value is weaker with the cross-validated ensemble sizes used in the main experiments.}
We set the sample size $(n_0, n_1)$ to $(500, 500)$, $(300, 700)$, $(200, 800)$, and $(100, 900)$, changing the degree of sample-size imbalance, and the results are shown in Figure~\ref{fig: 1d comparisons in appendix}.
We can see that the estimated log posterior odds from AdaBoost are overall shrunk toward 0, which is the initial value in the algorithm \citep{culp2007ada}, and this results in a positive bias in estimating the density ratios under the unbalanced settings.
\rnew{In contrast, for our proposed approach, although the magnitude of the log-density ratio is underestimated\rnew{, particularly with this small ensemble}, the sign of the estimate generally agrees with that of the truth.}
\rnew{The same tendency of the AdaBoost-based method to introduce bias is observed in the two-dimensional cases, as illustrated in Figures~\ref{fig:s3-global-shift-unbalanced}--\ref{fig:s3-local-dispersion-unbalanced}.}
}

\begin{figure}[htbp]
\centering
\includegraphics[width=\linewidth,height=0.76\textheight,keepaspectratio]{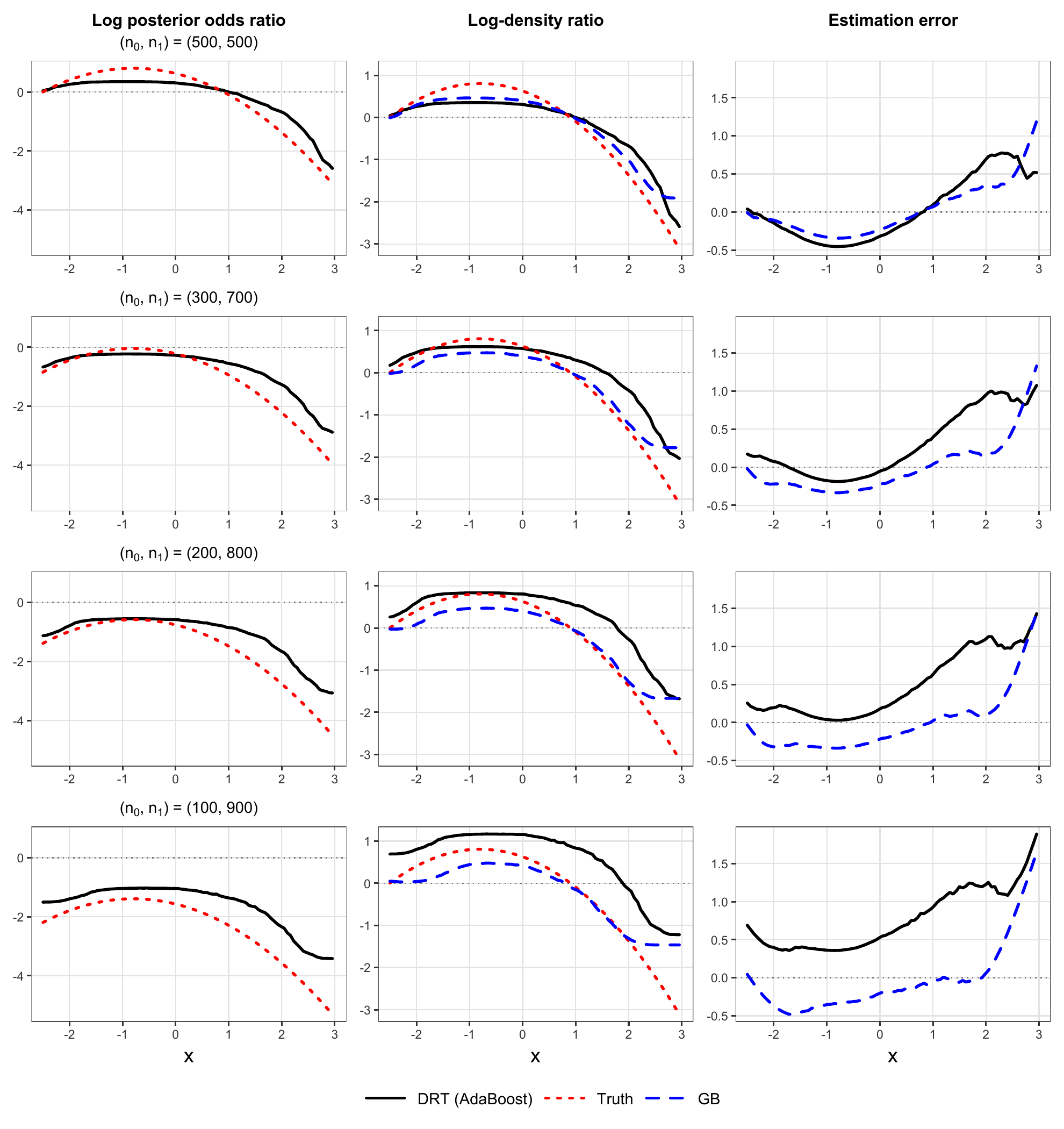}
\caption{\nanew{A comparison of the density-ratio trick (DRT) approach based on the AdaBoost and the proposed gradient boosting (GB) under the 1D example with different sample sizes.
The first column shows a comparison between the true log posterior odds and those estimated by AdaBoost, and the second column visualizes the log-density ratios estimated with DRT and GB along with the truth. 
Their estimation errors are visualized in the third column.
The estimated functions are the averages of the results based on 50 simulated data sets.
}}
\label{fig: 1d comparisons in appendix}
\end{figure}

\subsection{\rnew{Additional Results for the Two-dimensional Scenarios}}

\rnew{We provide additional visualizations corresponding to the settings described in Section~\ref{sec: 2d scenarios}.}
\rnew{Figures~\ref{figure: 2d visualization of the estimated ratios (appendix, balanced)} and~\ref{fig:s2-local-dispersion-balanced} show the estimated log-density ratios for the Global Shift and Local Dispersion scenarios, respectively, under the balanced sample-size setting. Figures~\ref{fig:s3-global-shift-unbalanced}--\ref{fig:s3-local-dispersion-unbalanced} show the Global Shift, Local Shift, and Local Dispersion scenarios, respectively, under the unbalanced sample-size setting.}
For each unbalanced scenario, the 95\% pointwise credible intervals exclude zero for more than 10\% of the observations, indicating posterior evidence of nonzero log-density ratios in the corresponding regions, although this pattern is not readily visible in the plots.

\begin{figure}[H]
\centering
\includegraphics[width=13.5cm]{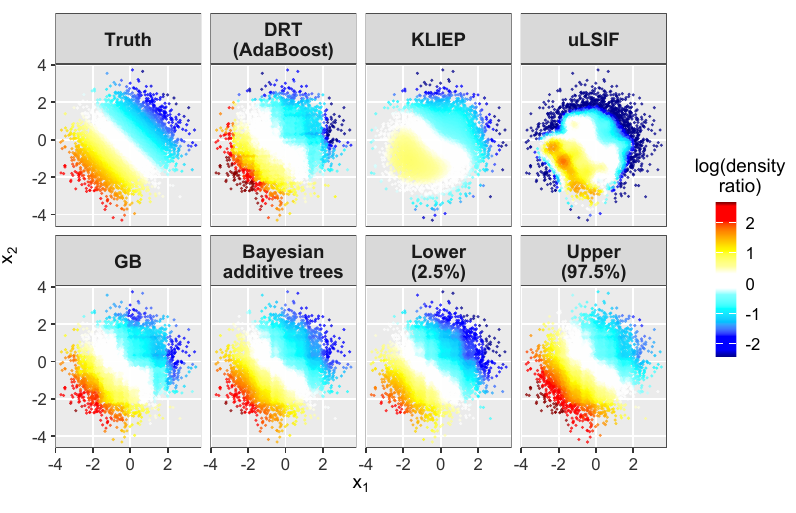}
\renewcommand{\thefigure}{S\arabic{figure}-1}
\caption{\rnew{Estimated log-density ratios for the two-dimensional Global Shift scenario ($n_0 = n_1 = 5000$). For Bayesian additive trees, the pointwise posterior mean and the 2.5\% and 97.5\% quantiles are shown. Because the results for FS and GB were similar, only GB is displayed.}}
\label{figure: 2d visualization of the estimated ratios (appendix, balanced)}
\end{figure}

\begin{figure}[H]
\ContinuedFloat
\centering
\includegraphics[width=13.5cm]{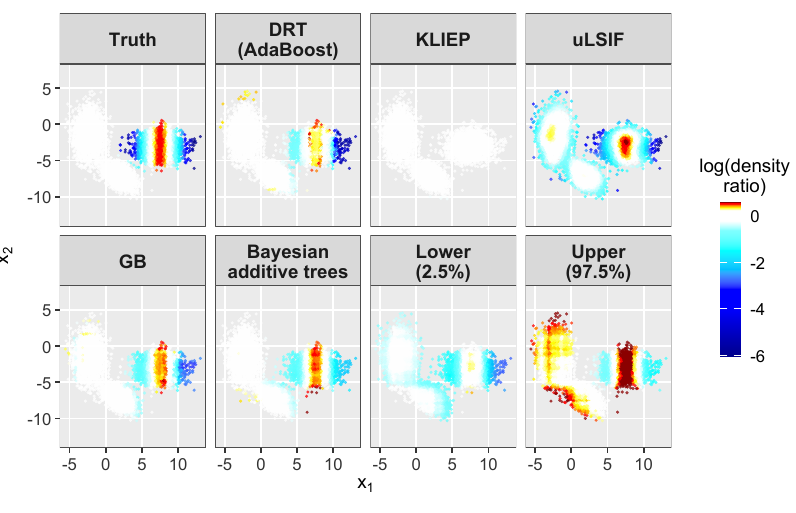}
\renewcommand{\thefigure}{S\arabic{figure}-2}
\caption{\rnew{Estimated log-density ratios for the two-dimensional Local Dispersion scenario ($n_0 = n_1 = 5000$). For Bayesian additive trees, the pointwise posterior mean and the 2.5\% and 97.5\% quantiles are shown. Because the results for FS and GB were similar, only GB is displayed.}}
\label{fig:s2-local-dispersion-balanced}
\end{figure}

\begin{figure}[H]
\centering
\includegraphics[width=13.5cm]{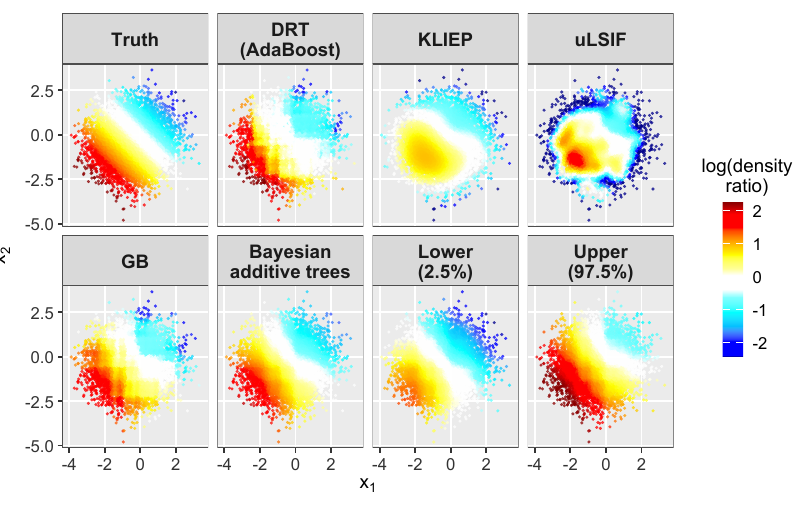}
\renewcommand{\thefigure}{S\arabic{figure}-1}
\caption{\rnew{Estimated log-density ratios for the two-dimensional Global Shift scenario ($n_0 = 9000$ and $n_1 = 1000$). For Bayesian additive trees, the pointwise posterior mean and the 2.5\% and 97.5\% quantiles are shown. Because the results for FS and GB were similar, only GB is displayed.}}
\label{fig:s3-global-shift-unbalanced}
\end{figure}

\begin{figure}[H]
\ContinuedFloat
\centering
\includegraphics[width=13.5cm]{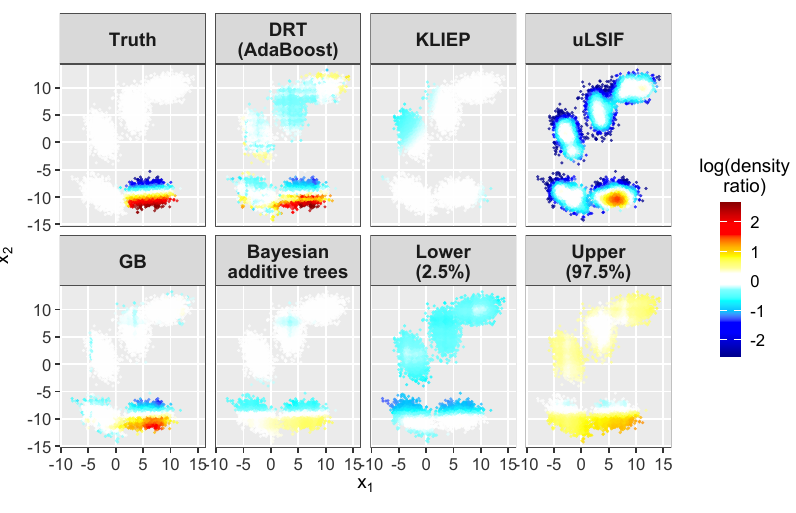}
\renewcommand{\thefigure}{S\arabic{figure}-2}
\caption{\rnew{Estimated log-density ratios for the two-dimensional Local Shift scenario ($n_0 = 9000$ and $n_1 = 1000$). For Bayesian additive trees, the pointwise posterior mean and the 2.5\% and 97.5\% quantiles are shown. Because the results for FS and GB were similar, only GB is displayed.}}
\label{fig:s3-local-shift-unbalanced}
\end{figure}

\begin{figure}[H]
\ContinuedFloat
\centering
\includegraphics[width=13.5cm]{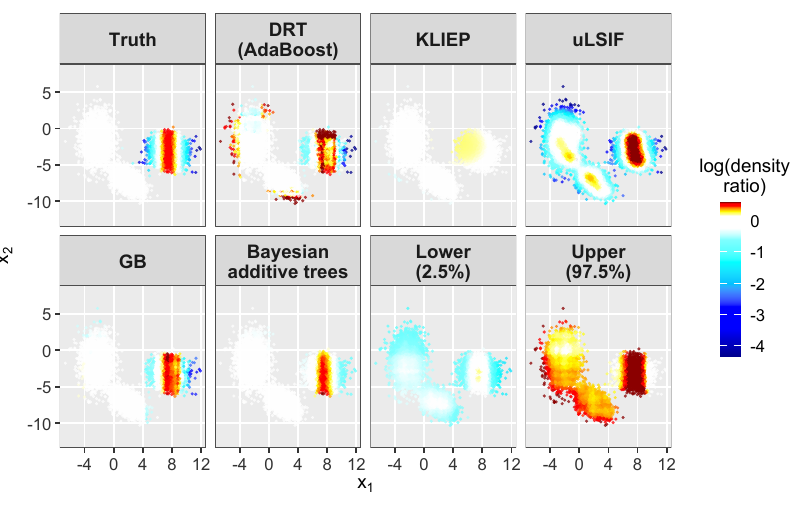}
\renewcommand{\thefigure}{S\arabic{figure}-3}
\caption{\rnew{Estimated log-density ratios for the two-dimensional Local Dispersion scenario ($n_0 = 9000$ and $n_1 = 1000$). For Bayesian additive trees, the pointwise posterior mean and the 2.5\% and 97.5\% quantiles are shown. Because the results for FS and GB were similar, only GB is displayed.}}
\label{fig:s3-local-dispersion-unbalanced}
\end{figure}

\begin{figure}[H]
\centering
\begin{tabular}{c}
   \includegraphics[width=10.5cm]{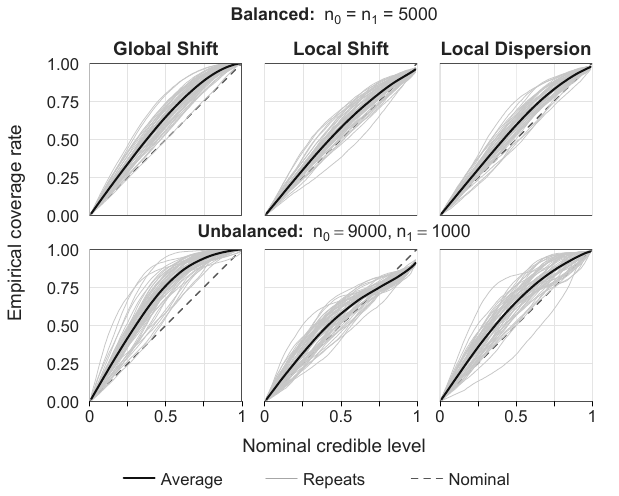}
\end{tabular}
\caption{\nanew{The calibration plots for the two-dimensional examples, in which the posterior uncertainty and the nominal coverage rates (namely, the ratio of observations included in the pointwise credible intervals) are compared under different probabilities. 
Each gray curve corresponds to one of the 50 simulated data sets, and the black curve indicates their averages.
}}
\label{figure: coverage rate 2D}
\end{figure}

\nanew{
    \rnew{We also evaluate the frequentist coverage rates of the credible intervals using the calibration plots shown in Figure~\ref{figure: coverage rate 2D}.}
    We can observe only small over-coverage in the balanced scenarios, in which the data sets provide more information about the two-sample difference.
    We therefore expect that the over-coverage and under-coverage under the unbalanced sample sizes are due to the structures of the differences that are difficult to accurately estimate with the small amount of information. 
    For example, the diagonal shift in the Global Shift scenario is hard to describe with the axis-aligned tree partitions, and, for the Local Shift scenario, there is a sharp change in the density ratios in the local subset, but only a small number of observations from the second group (the smaller group) are observed there.
    Again, we note that such differences are accurately estimated under the balanced sample sizes in which richer information about the difference is provided from the data.
    }

\subsection{\rnew{Numerical Diagnostics for the Calibrated Discriminative Classifier}}
\label{sec:cdc-calibration}

\rnew{The analysis in Section~\ref{sec: 2d scenarios} uses the Calibrated Discriminative Classifier (CDC) \citep{cranmer2015approximating}, applied to the AdaBoost classification scores. Let $s_i$ denote the AdaBoost score after subtracting the empirical log class-odds offset $\log(n_1/n_0)$. The submitted implementation fits separate one-dimensional Gaussian kernel density estimates, $\widehat g_0$ and $\widehat g_1$, to the scores from groups 0 and 1. It then estimates the log-density ratio at observation $i$ as $\log\{\widehat g_0(s_i)/\widehat g_1(s_i)\}$, using linearly interpolated kernel density values. At extreme scores, numerical underflow can set an interpolated density value to zero and thereby produce a non-finite log-density ratio. The main-text table therefore reports finite-only summaries for CDC in the affected cells.}

\rnew{As a numerical sensitivity analysis, the stable implementation uses the same AdaBoost scores, group-specific score samples, and Gaussian-kernel bandwidths. It first computes $\log\widehat g_0(s_i)-\log\widehat g_1(s_i)$ instead of taking the logarithm after forming the density ratio. If an interpolated density is still numerically zero, the implementation evaluates each Gaussian kernel density directly in log space using the log-sum-exp identity and then takes the difference. Thus, the stable calculation changes the numerical evaluation of the same CDC estimator without refitting AdaBoost or changing the inferential target. Table~\ref{table: cdc numerical diagnostics} shows that this calculation removes non-finite values but does not improve accuracy in the affected scenarios, where large tail errors remain.}

\begin{table}[H]
\centering
\caption{\rnew{Numerical diagnostics for CDC in the two-dimensional cells with non-finite estimates.}}
\label{table: cdc numerical diagnostics}
\begin{tabular}{lccc}
\toprule
\rnew{Scenario} & \rnew{$n_0/(n_0+n_1)$} & \rnew{Finite repetitions} & \rnew{Stable MSE (SE)} \\
\midrule
\rnew{Local Shift} & \rnew{0.5} & \rnew{24/50} & \rnew{18.489 (10.053)} \\
\rnew{Local Shift} & \rnew{0.9} & \rnew{46/50} & \rnew{1.813 (0.215)} \\
\rnew{Local Dispersion} & \rnew{0.5} & \rnew{21/50} & \rnew{50.149 (27.515)} \\
\rnew{Local Dispersion} & \rnew{0.9} & \rnew{40/50} & \rnew{8.068 (4.243)} \\
\bottomrule
\end{tabular}
\begin{tablenotes}
\small
\item \rnew{Finite repetitions refer to the submitted CDC implementation. Stable MSEs use all 50 repetitions and are computed using the log-space evaluation described above.}
\end{tablenotes}
\end{table}

\subsection{\rnew{Illustrative Density-Ratio Estimates in the 20-dimensional Location Shift Scenario}}

\begin{figure}[H]
\centering
\includegraphics[width=13.5cm]{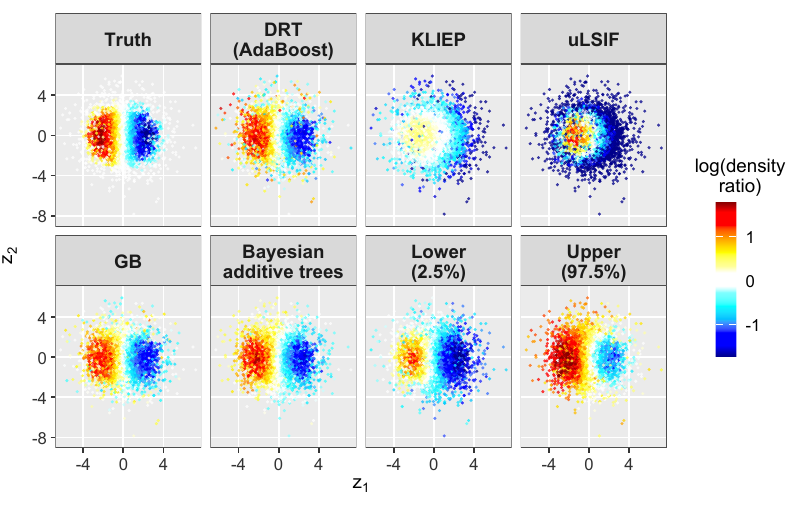}
\caption{\rnew{Illustrative comparison of the true and estimated log-density ratios projected onto the first two latent coordinates under the 20-dimensional Location Shift scenario, with $n_0=n_1=5000$. For the Bayesian additive tree model, the pointwise posterior mean and the 2.5\% and 97.5\% quantiles are shown. Because the results for FS and GB were similar, only GB is displayed.}}
\label{fig:supp_20d_location_density_ratio}
\end{figure}

\rnew{To illustrate the spatial behavior of the density-ratio estimators, we present results for the Location Shift scenario with balanced sample sizes in Figure~\ref{fig:supp_20d_location_density_ratio}. This scenario permits a direct visualization of the true and estimated log-density ratios in the first two latent coordinates. Quantitative comparisons across the 20-dimensional scenarios are reported in Section~\ref{sec: higher-dimensional settings}.}

\begin{figure}[H]
\centering
\includegraphics[width=11.8cm]{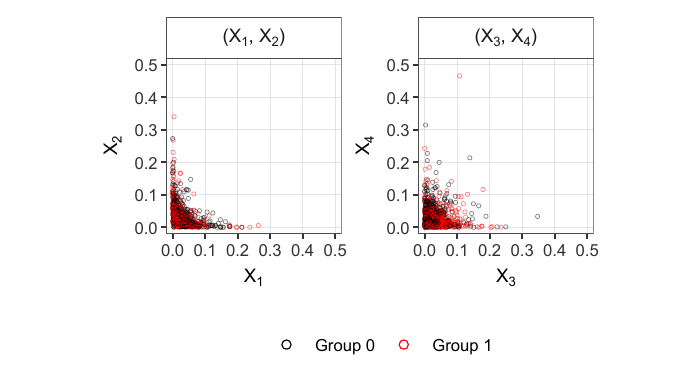}
\caption{\rnew{Pairwise projections onto $(X_1, X_2)$ and $(X_3, X_4)$ of a simulated data set under the transformed 20-dimensional location-shift scenario. Black and red points represent observations from groups 0 and 1, respectively.}}
\label{figure: alt 20d scenario}
\end{figure}

\nanew{
\subsection{\rnew{Alternative Simulation Scenarios with 20-dimensional Structures}}
    \rnew{In this section, we modify the 20-dimensional simulation scenarios so that the data structure resembles that of the real microbiome data set analyzed in Section~\ref{sec:real_data}.}
    To this end, we employ the following procedure:
    \begin{enumerate}
        \item \rnew{We simulate a data set under the Global Shift, Location Shift, or Dispersion Difference scenario.}
        \item We transform their marginal distributions by applying the CDFs of the Gaussian distributions
        \footnote{\rnew{
        For the Location Shift and Dispersion Difference scenarios, the reference Gaussian parameters are selected to match the moments of the mixture component with the larger variances, $N(\mu_2,\Sigma_2)$. For Global Shift, the reference Gaussian distribution is centered at zero and has the marginal variances implied by the projection and noise model.
        }
        }
        and then the inverse CDF of the beta distribution $\mathrm{Beta}(0.5, 10)$.
    \end{enumerate}
    As a result, the marginal distributions are right-skewed and have a large probability around 0, and the distributional structure is similar to the high-dimensional abundance data, as illustrated in Figure~\ref{figure: alt 20d scenario}.
    \rnew{The estimation results based on 50 simulated data sets are summarized in Table~\ref{table: MSE(multi, alt)}. Under the unbalanced sample sizes, BAT has the lowest MSE in the Global Shift scenario, while GB has the lowest MSE in the Location Shift and Dispersion Difference scenarios, and both GB and FS have lower MSE than DRT in all three scenarios. Under the balanced sample sizes, GB has lower MSE than DRT in the Global Shift and Location Shift scenarios, whereas DRT has slightly lower MSE than GB in the Dispersion Difference scenario.}
}

\begin{table}[H]
\centering
\caption{A comparison of the MSEs and their standard errors\\
in the alternative 20-dimensional scenarios.
}\label{table: MSE(multi, alt)}
\setlength{\tabcolsep}{9pt}%
\begin{tabular}{lcccccc}\toprule
    &\multicolumn{2}{c}{\rnew{\textbf{Global Shift}}}
    &\multicolumn{2}{c}{\textbf{Location Shift}}
    &\multicolumn{2}{c}{\textbf{Dispersion}}
    \\
    \cmidrule(r){2-3}\cmidrule(r){4-5}\cmidrule(r){6-7}
    $n_0 / (n_0 + n_1)$ &\rnew{0.5}&\rnew{0.9}&0.5&0.9&0.5&0.9\\\midrule
GB &\rnew{0.057} &\rnew{0.099} &\rnew{0.061} &\rnew{0.101} &\rnew{0.140} &\rnew{0.222} \\
&\rnew{(0.001)} &\rnew{(0.002)} &\rnew{(0.001)} &\rnew{(0.002)} &\rnew{(0.003)} &\rnew{(0.005)} \\
FS &\rnew{0.057} &\rnew{0.105} &\rnew{0.064} &\rnew{0.107} &\rnew{0.150} &\rnew{0.228} \\
&\rnew{(0.001)} &\rnew{(0.003)} &\rnew{(0.001)} &\rnew{(0.002)} &\rnew{(0.004)} &\rnew{(0.006)} \\
BAT &\rnew{0.038} &\rnew{0.079} &\rnew{0.062} &\rnew{0.108} &\rnew{0.179} &\rnew{0.363} \\
&\rnew{(0.001)} &\rnew{(0.002)} &\rnew{(0.001)} &\rnew{(0.002)} &\rnew{(0.004)} &\rnew{(0.008)} \\
DRT (AdaBoost) &\rnew{0.061} &\rnew{0.139} &\rnew{0.070} &\rnew{0.173} &\rnew{0.137} &\rnew{0.265} \\
&\rnew{(0.002)} &\rnew{(0.003)} &\rnew{(0.001)} &\rnew{(0.003)} &\rnew{(0.003)} &\rnew{(0.005)} \\
CDC (AdaBoost) &\rnew{0.489} &\rnew{1.211} &\rnew{0.160} &\rnew{0.600} &\rnew{0.544} &\rnew{1.729} \\
&\rnew{(0.023)} &\rnew{(0.045)} &\rnew{(0.003)} &\rnew{(0.020)} &\rnew{(0.023)} &\rnew{(0.055)} \\
KLIEP &\rnew{0.935} &\rnew{0.931} &\rnew{0.684} &\rnew{0.683} &\rnew{0.538} &\rnew{0.539} \\
&\rnew{(0.012)} &\rnew{(0.013)} &\rnew{(0.002)} &\rnew{(0.003)} &\rnew{(0.001)} &\rnew{(0.001)} \\
uLSIF &\rnew{3.020} &\rnew{2.019} &\rnew{20.022} &\rnew{14.810} &\rnew{0.538} &\rnew{0.539} \\
&\rnew{(0.339)} &\rnew{(0.318)} &\rnew{(8.211)} &\rnew{(7.600)} &\rnew{(0.001)} &\rnew{(0.001)} \\

            \bottomrule
\end{tabular}%
\begin{tablenotes}
\small
\item Notes: The methods are GB (gradient boosting), FS (forward stagewise), BAT (Bayesian additive trees), DRT \nanew{and CDC} (density-ratio trick and \nanew{calibrated discriminative classifier} based on AdaBoost), and the two kernel-based methods KLIEP and uLSIF.
\end{tablenotes}
\end{table} 

\clearpage
\subsection{\rnew{Additional One-dimensional Calibration Results}}
\begin{figure}[htbp]
\centering
\includegraphics[width=\linewidth]{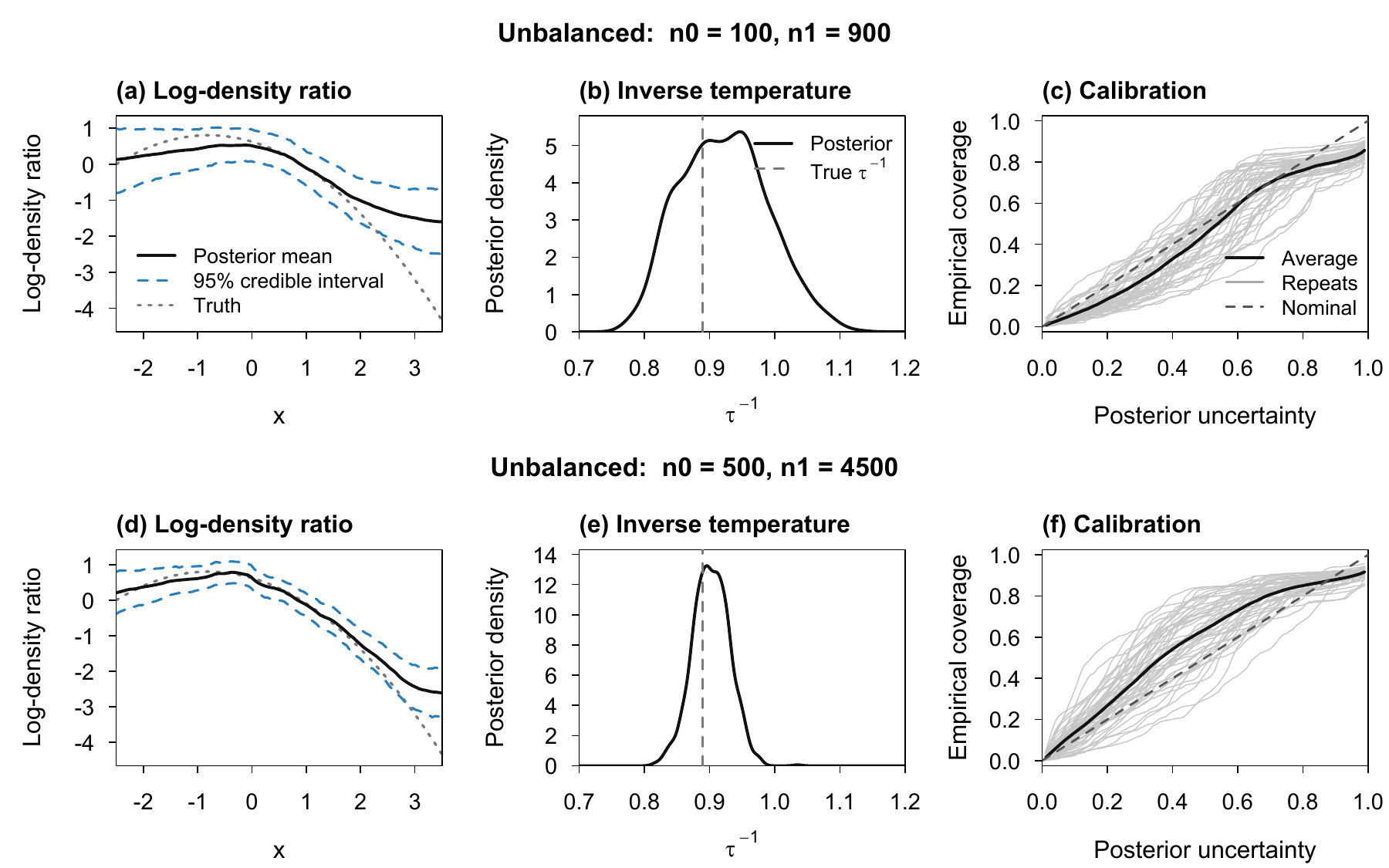}
\caption{\rnew{An evaluation of posterior distributions under the two unbalanced 1D sample-size settings, $(n_0,n_1)=(100,900)$ and $(500,4500)$. Each left plot compares the true log-density ratio with the pointwise posterior mean and 95\% credible interval. Each middle plot shows the posterior distribution of the inverse temperature $\tau^{-1}$ and the true Bhattacharyya coefficient. Each right plot compares the nominal credible level with empirical coverage over 50 simulation repeats.}}
\label{fig: 1D illustrations unbalanced}
\end{figure}

\rnew{Figure~\ref{fig: 1D illustrations unbalanced} presents the posterior distributions and calibration results for the unbalanced one-dimensional settings.}

\clearpage
\section{\rnew{Additional Real Data Analysis Results}}\label{supp_sec:RDA}

\begin{sloppypar}
\rnew{The implementations of ICFM and MB-GAN used in Section~\ref{sec:real_data} are available at \url{https://github.com/atong01/conditional-flow-matching} and \url{https://github.com/zhanxw/MB-GAN}, respectively.}
\end{sloppypar}

\begin{figure}[ht!]
    \centering
    \includegraphics[width=\linewidth]{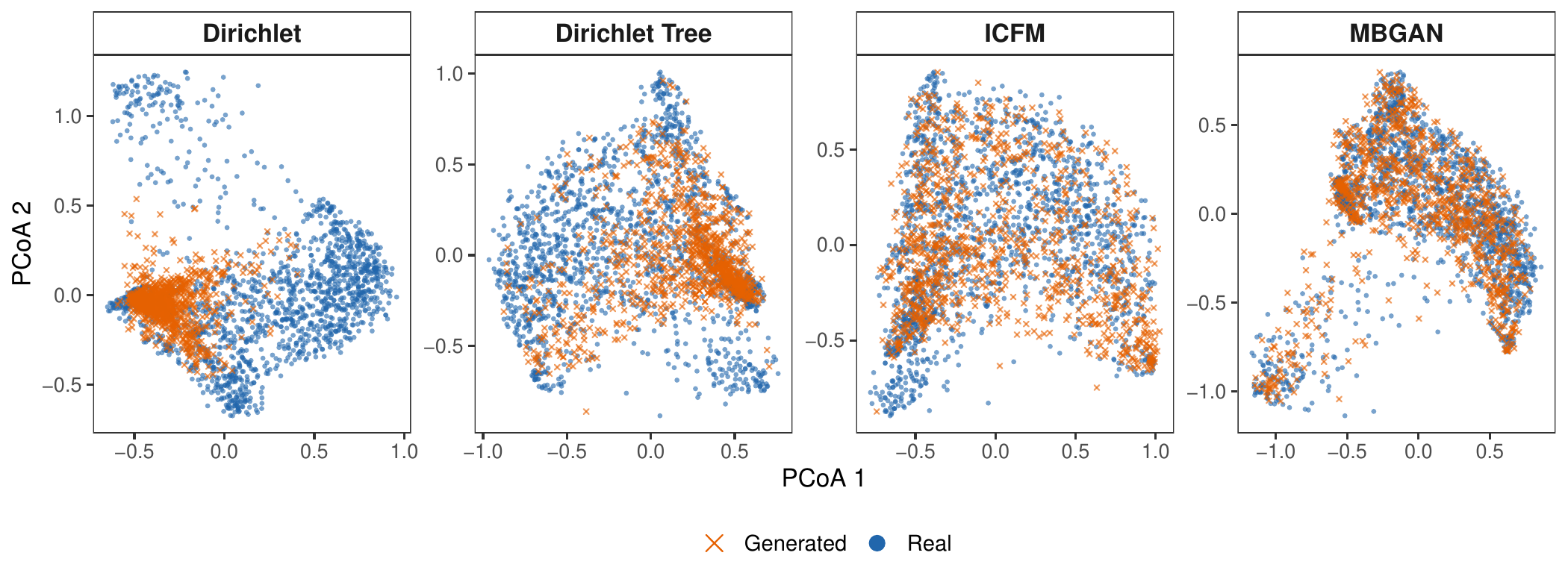}
    \caption{\yxnew{Principal Coordinates Analysis (PCoA) plots based on Bray-Curtis distances among training sample observations.
    The blue and orange points represent the real sample and the generated sample, respectively. }
   }
    \label{fig:PCoA_raw_train}
\end{figure}

\yxnew{Figure~\ref{fig:PCoA_raw_train} is the training sample equivalent to Figure~\ref{fig:PCoA_raw}. The takeaway is similar to Figure~\ref{fig:PCoA_raw}: MB-GAN produces the most overlapping samples visualized on the PCoA plot, while other methods present large non-overlapping synthetic samples compared to the observed real samples.}

\newpage 

\begin{figure}[H]
  \centering
  \includegraphics[width=\linewidth]{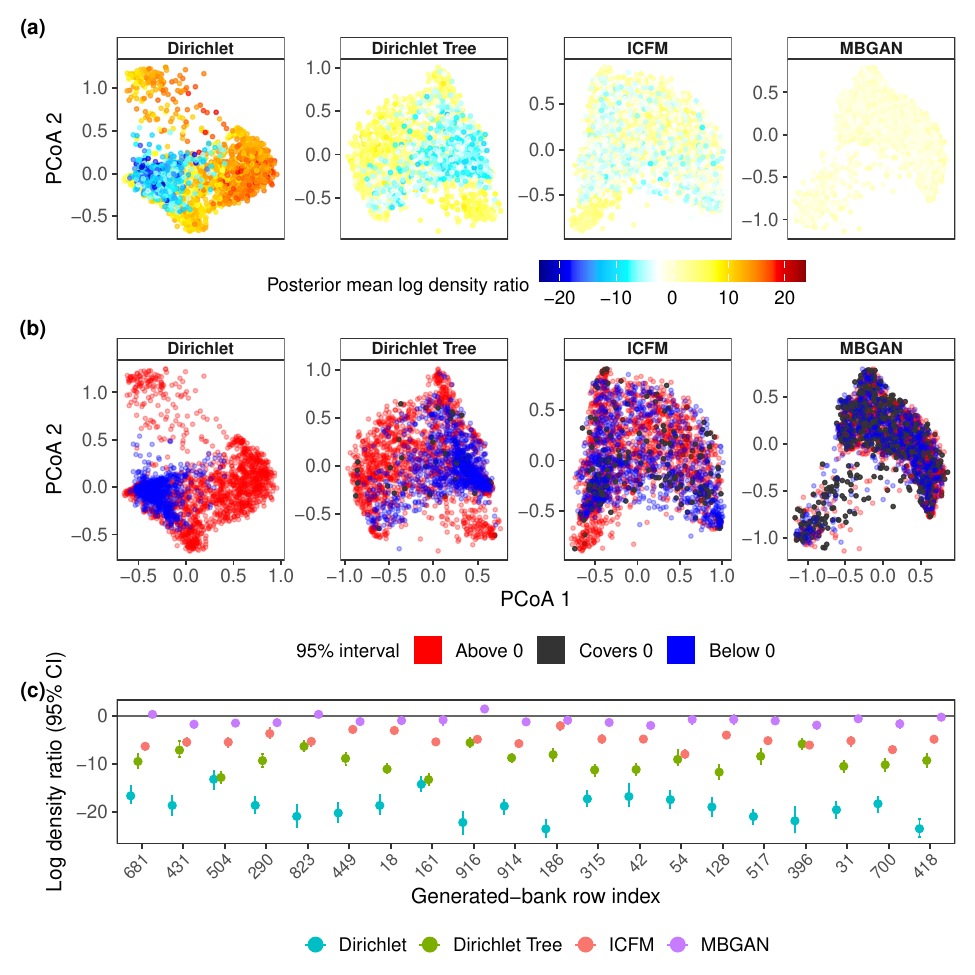}
  \caption{\yxnew{For the training data: (a) PCoA plots (x-axis and y-axis are the first and second principal components) colored by the density ratio estimates of the training sample density over the generated sample density, and (b) whether the 2.5\% and 97.5\% credible bands cover 0; (c) For the top 20 observations with the largest absolute log-density ratio averaged over all methods, the point estimates and the credible intervals for each generative model.}}
  \label{fig:RDA_DRE_train}
\end{figure}

Figure~\ref{fig:RDA_DRE_train} provides the training sample equivalent to Figure~\ref{fig:RDA_DRE_test}.

\newpage

\begin{figure}[H]
    \centering
    \includegraphics[width=\linewidth]{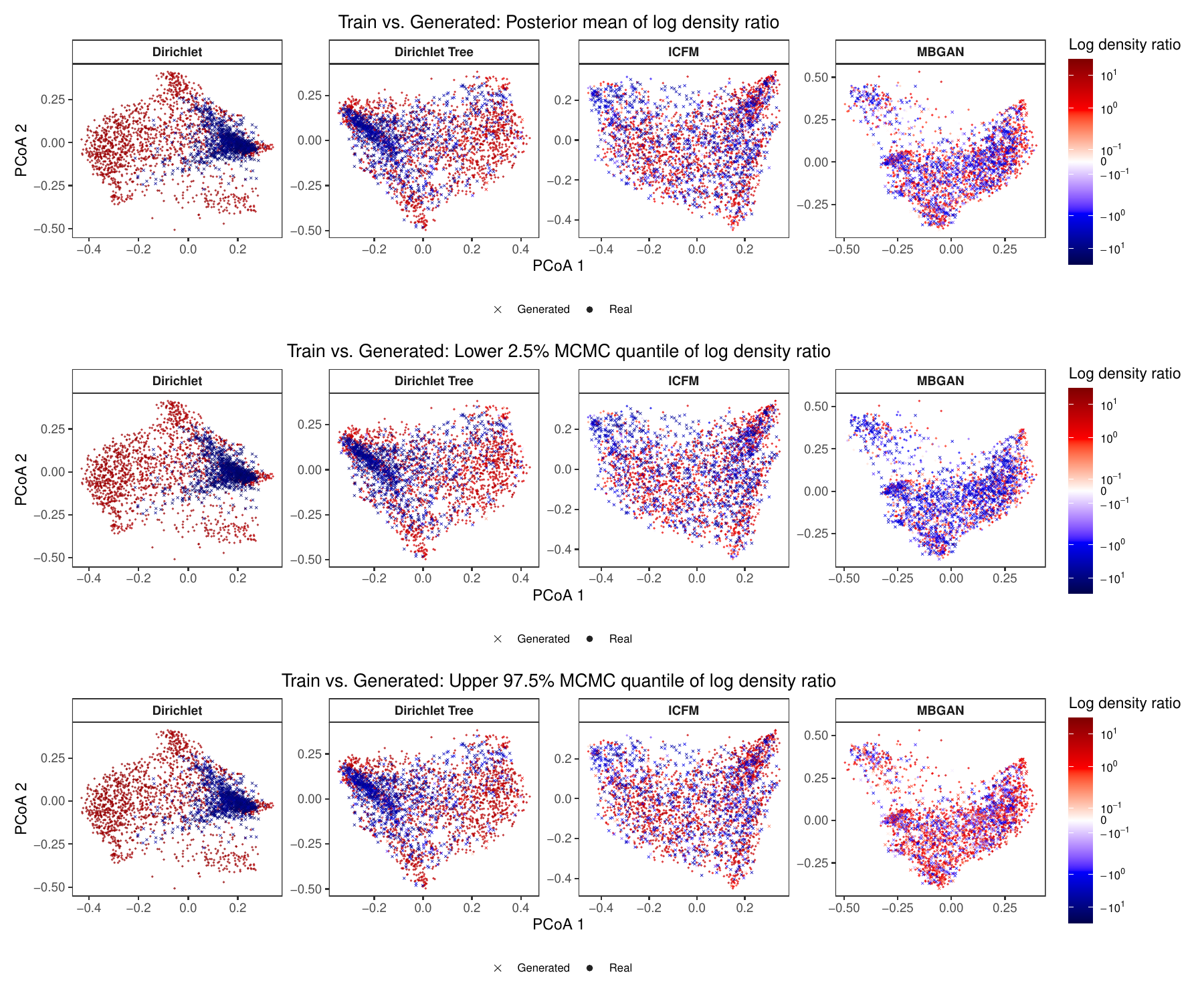}
    \caption{PCoA plots colored by the density ratio estimates of the training sample density over the generated sample density, along with the 2.5\% and 97.5\% credible bands.}
    \label{fig:supp_PCoA_DRE_train}
\end{figure}

\begin{figure}[H]
    \centering
    \includegraphics[width=\linewidth]{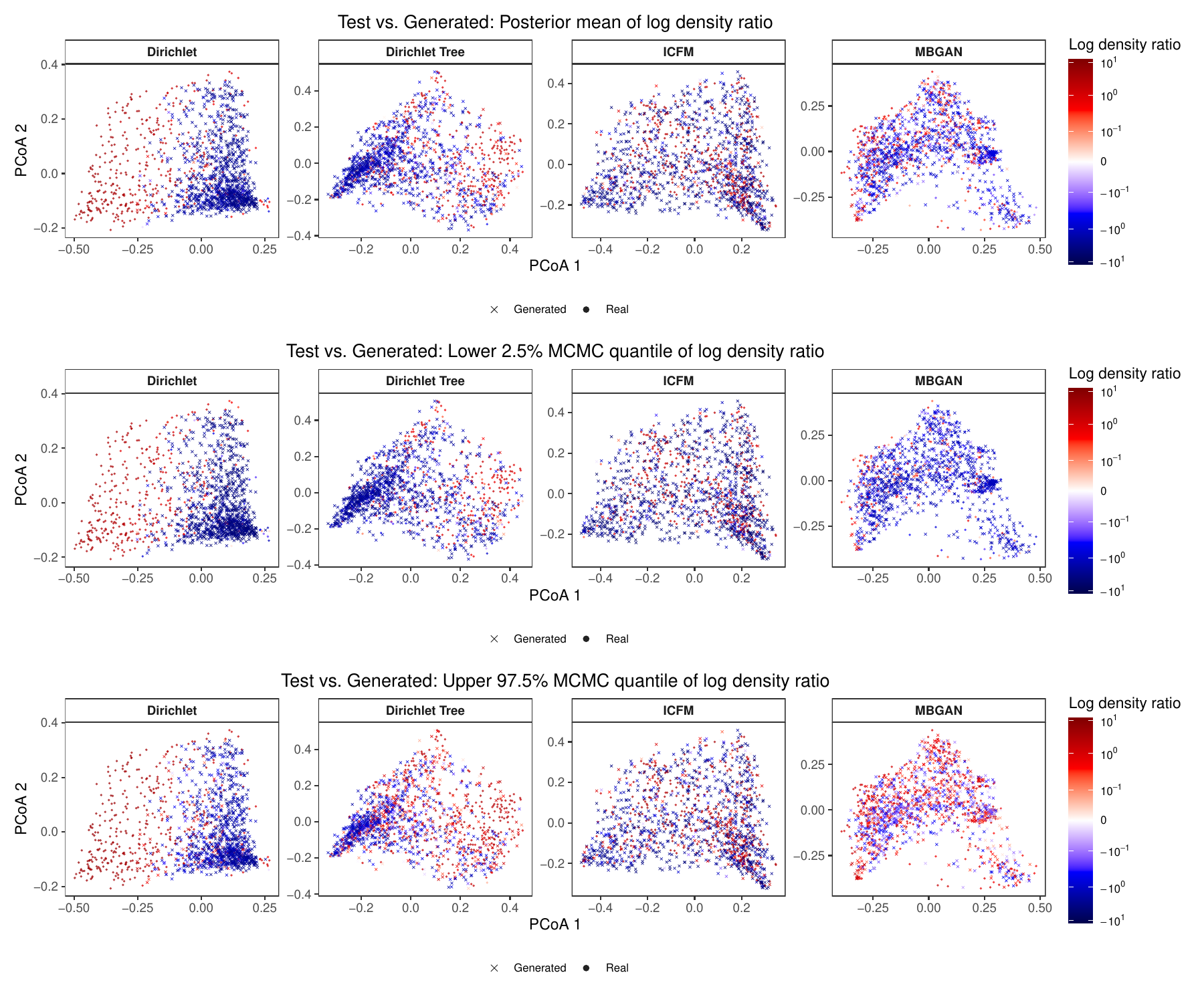}
    \caption{PCoA plots colored by the density ratio estimates of the testing sample density over the generated sample density, and the 2.5\% and 97.5\% credible bands.}
    \label{fig:supp_PCoA_DRE_test}
\end{figure}

\rnew{Figures~\ref{fig:supp_PCoA_DRE_train} and \ref{fig:supp_PCoA_DRE_test} provide additional real-data results comparing credible intervals across the generative models. The PCoA plots show the credible bands for all samples.}

\begin{figure}
    \centering
    \includegraphics[width=\linewidth]{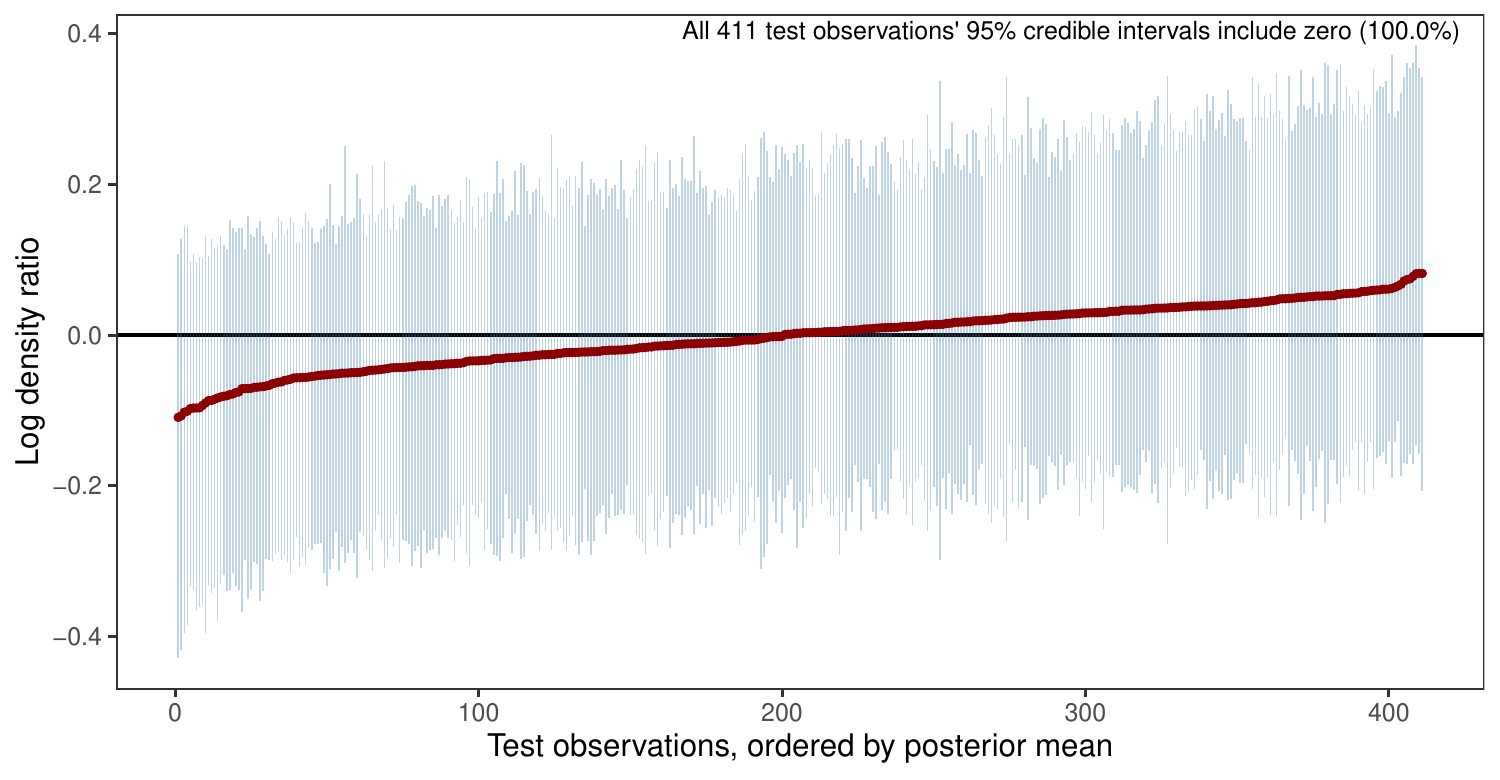}
    \caption{\yxnew{Null-experiment credible bands. Pointwise posterior means and 95\% credible intervals for the log-density ratio evaluated on the 411 held-out test observations, ordered by posterior mean. The horizontal line marks the null value of zero. All 411 credible intervals contain zero, corresponding to 100\% observed coverage under the null experiment.}}
    \label{fig:null_test}
\end{figure}

\yxnew{\rnew{As mentioned at the end of Section~\ref{sec:real_data}, the null experiment serves as a check for potential overfitting in the high-dimensional microbiome data. Figure~\ref{fig:null_test} shows that the credible intervals for all test observations contain zero. Compared with the real-data results in Figure~\ref{fig:RDA_DRE_test}, this finding indicates that the credible intervals excluding zero in the real-data analysis are unlikely to result from overfitting.}}

\clearpage
\bibliographystyle{agsm} 
\bibliography{references} 

\end{document}

\end{document}